\pdfoutput=1
\documentclass[journal]{IEEEtran}
\usepackage{fix-cm}
\usepackage{cite}
\usepackage{amsmath,amssymb,amsfonts}
\usepackage{algorithm}
\usepackage{algpseudocode}
\usepackage[utf8]{inputenc}
\usepackage{textgreek}
\usepackage[caption=false,font=footnotesize]{subfig}
\usepackage{tabularx}
\usepackage{array}
\usepackage{bm}
\usepackage{float}
\usepackage{booktabs}
\usepackage{multirow}
\usepackage{acronym}
\input{acronym}
\usepackage{graphicx,color}
\usepackage{textcomp}
\usepackage{xcolor}
\usepackage{microtype}
\usepackage{tikz}
\usepackage{hyperref}
\hypersetup{hidelinks}
\newcommand{\manutablefont}{\fontsize{8}{9.5}\selectfont}
\newcolumntype{Y}{>{\centering\arraybackslash\leavevmode}X}

\def\BibTeX{{\rm B\kern-.05em{\sc i\kern-.025em b}\kern-.08em
    T\kern-.1667em\lower.7ex\hbox{E}\kern-.125emX}}

\newcommand{\T}{\mathsf{T}}

\begin{document}


\title{Orientation-Aware Deep Learning Framework for RSSI-Based Indoor Localization in RIS-Aided mmWave Systems}

\author{Varun~S.~Advani,~\IEEEmembership{Graduate Student Member,~IEEE,}
Ahmed~Nasser,~\IEEEmembership{Senior Member,~IEEE,}
Mohamed~Y.~Selim,~\IEEEmembership{Senior Member,~IEEE,}
and~Ahmed~M.~Eltawil,~\IEEEmembership{Senior Member,~IEEE}%
\thanks{V. S. Advani and M. Y. Selim are with the Department of Electrical and Computer Engineering, Iowa State University of Science and Technology, USA (e-mail: vsadvani@iastate.edu; myoussef@iastate.edu)}%
\thanks{A. Nasser and A. M. Eltawil are with the Computer, Electrical, and Mathematical Sciences and Engineering Division, King Abdullah University of Science and Technology (KAUST), Thuwal, Saudi Arabia (e-mail: ahmednasser.ismail@kaust.edu.sa; ahmed.eltawil@kaust.edu.sa)}}

\maketitle

\begin{abstract}
\Ac{RIS}-aided \ac{mmWave} systems provide directional measurements of received power for indoor localization, but device orientation changes the fingerprints. This work proposes an \ac{ODSG} deep learning framework for \ac{RSSI}-based indoor localization. An orientation classifier estimates device orientation from \ac{RSSI}, {so inference needs only received power}. An embedding of the predicted angle and posterior entropy controls feature modulation and soft expert gating. Parallel convolutional and statistical streams are fused through \ac{BiLSTM} processing and temporal attention. Single-user measurements {in an indoor laboratory} at 28 GHz under \ac{NLoS} conditions yield a median localization error of 0.141 m, a 25.8\% improvement over the closest evaluated baseline {(supplied with the measured orientation)}, and 92.3\% orientation classification accuracy across four discrete classes. {In spatial regions withheld from training, \ac{ODSG} has an average median error of 0.247 m, against 0.392 m for a validation-selected sequence model.} {A separate offline evaluation constructs \ac{NOMA}-style power mixtures from the single-user measurements and measures localization accuracy and computational scaling with up to eight emulated users whose power contributions overlap, under the specified mixture model.}

\end{abstract}

\begin{IEEEkeywords}
Indoor localization, reconfigurable intelligent surface, millimeter wave, deep learning, orientation prediction, multi-user systems, received signal strength indicator, mixture of experts.
\end{IEEEkeywords}

\acresetall

\section{Introduction}

\Ac{RF}-based indoor localization is a component of \ac{ISAC} for \ac{6G} wireless systems~\cite{liu2022isac_survey}. Applications include smart building and industrial automation, asset tracking, autonomous robotic navigation, and augmented reality, which require accurate, timely position estimates~\cite{li2024indoor}. Unlike outdoor positioning supported by \ac{GNSS}, indoor positioning must contend with strong signal attenuation, multipath propagation, and frequent \ac{NLoS} conditions~\cite{zafari2019indoor_survey}.

\begin{table*}[!t]
\centering
\caption{Representative Localization and Adaptive RIS-Control Approaches}
\label{tab:related_work_comparison}
\fontsize{8}{9.6}\selectfont
\setlength{\tabcolsep}{3pt}
\begin{tabularx}{\textwidth}{@{}>{\raggedright\arraybackslash\leavevmode}p{0.17\textwidth}>{\raggedright\arraybackslash\leavevmode}p{0.18\textwidth}*{5}{Y}@{}}
\toprule
\textbf{Reference} & \textbf{Approach} & {\textbf{RSS-Based}} & {\textbf{Orientation Handling}} & \textbf{Multi-User} & \textbf{DL-Based} & \textbf{Experimental} \\
\midrule
Huang et al.~\cite{huang2023joint} & Geometry & -- & -- & -- & -- & {--} \\
Sun et al.~\cite{sun2024near} & Geometry & -- & -- & -- & -- & {--} \\
Dardari et al.~\cite{dardari2021nlos} & Geometry & -- & -- & -- & -- & -- \\
Zhang et al.~\cite{zhang2021metalocalization} & {RSS+optimization} & {\checkmark} & -- & \checkmark & {--} & -- \\
Keykhosravi et al.~\cite{keykhosravi2022ris} & Geometry & -- & -- & -- & -- & -- \\
Rizk et al.~\cite{rizk2022robust} & RSSI+Fusion & \checkmark & -- & -- & \checkmark & \checkmark \\
Rahal et al.~\cite{rahal2023ris} & {Channel metrics+LS} & {Gain-derived AoD} & -- & -- & -- & \checkmark \\
Aguiar et al.~\cite{aguiar2024deep} & {Complex-signal DL} & {--} & -- & -- & \checkmark & {--} \\
Nguyen et al.~\cite{nguyen2010reconfigurable} & {RSSI matching} & \checkmark & -- & -- & {--} & -- \\
Advani et al.~\cite{advani2025dsog} & RSSI+DL & \checkmark & {Known} & -- & \checkmark & \checkmark \\
{Chou et al.~\cite{chou2025drl}} & {DDPG precoding} & {--} & {N/A} & {Sim. comm.} & {\checkmark} & {--} \\
\midrule
\textbf{This Work} & \textbf{RSSI+DL} & \checkmark & {Inferred} & {Emulated} & \checkmark & {\checkmark} \\
\bottomrule
\end{tabularx}
\par\smallskip
\begin{minipage}{\textwidth}
\fontsize{7.5}{9}\selectfont\raggedright
\checkmark: addressed; --: not addressed; N/A: not a localization task; DL: deep learning; LS: least squares.\\
RSS: received signal strength; AoD: angle of departure.\\
{RSS-Based: inference uses received strengths, possibly with other modalities; Experimental: physical measurements.}\\
{Sim. comm.: simulated multiuser communication; Emulated: offline mixtures of measured single-user RSSI.}\\
{Chou et al. address adaptive communication control and provide methodological context for learning in RIS systems.}
\end{minipage}
\end{table*}

\Ac{mmWave} systems offer large bandwidth and compact directional antenna arrays for indoor localization~\cite{busari2018mmwave_survey}. Large bandwidth improves the temporal resolution of time-based measurements, while directional beams provide spatial information~\cite{shahmansoori2018posnavmmwave}. The present work exploits received power across beam and \acs{RIS} configurations. Blockage, attenuation over distance, and limited diffraction can create indoor coverage gaps~\cite{palacios2019mmwave}.

\Ac{RIS} technology uses programmable reflection coefficients to shape propagation paths and mitigate blockage~\cite{liu2021reconfigurable}. {The available phase shifts and reflection amplitudes depend on the hardware and may be coupled~\cite{abeywickrama2020practical,chou2025drl}.} Configuring the elements can create \ac{VLoS} paths and provide additional spatial diversity for localization under \ac{NLoS} conditions~\cite{rahal2023ris}. Channel-based configuration methods often require \ac{CSI}, whose acquisition is challenging for passive surfaces~\cite{11218162}.

Predefined beamforming codebooks allow \ac{RSSI} sweeps to be collected without explicit \ac{CSI} acquisition~\cite{11175018}. Common wireless hardware provides received signal strength measurements without coherent phase estimation, although fluctuations and multipath complicate the mapping from power to position~\cite{rizk2022robust}. By varying the reflected paths, \ac{RIS} configurations can produce spatially distinguishable fingerprints~\cite{nguyen2010reconfigurable}. Deep learning can model the nonlinear mapping from \ac{RSSI} to position~\cite{burghal2020comprehensive}. Related approaches include recurrent estimators and methods that combine learning with geometric estimation~\cite{aguiar2024deep,fadakar2024multi}.

\subsection{Related Works}

Localization in \ac{mmWave} systems without \ac{RIS} is well studied. Shahmansoori et al.~\cite{shahmansoori2018posnavmmwave} jointly estimate position and orientation from array observations and analyze direct and reflected propagation. Such approaches illustrate how resolvable multipath can supply geometric information. RIS-assisted methods add controllable reflected paths, whose usefulness depends on the available measurements and geometry.

Work on \ac{RIS}-aided indoor localization includes geometric estimation, fingerprinting, and learned estimators (Table~\ref{tab:related_work_comparison}). Geometry-based approaches extract position information from channel parameters or known array geometry~\cite{huang2023joint,sun2024near,dardari2021nlos,keykhosravi2022ris}. Rahal et al.~\cite{rahal2023ris} combine channel parameters obtained from indoor channel sounding with least-squares positioning. Zhang et al.~\cite{zhang2021metalocalization} instead optimize RIS configurations and RSS-based location decision functions for multiuser localization.

\looseness=1\relax {Nguyen et al.~\cite{nguyen2010reconfigurable} study RIS state selection and weighted-neighbor fingerprinting on simulated radio maps. Aguiar et al.~\cite{aguiar2024deep} study recurrent and hybrid learned localization using simulated complex received signals and RIS coefficients, including phase information. Section~\ref{sec:experimental_setup} evaluates Random Forest and neural-network fingerprint estimators of the kinds compared by Nguyen et al., together with recurrent, attention, and convolutional sequence models, on the measured received power.} {Our prior \ac{DSOG} framework~\cite{advani2025dsog} combined orientation-gated experts, a statistical stream, and \ac{BiLSTM} processing, with orientation supplied to the localization model. The present work estimates orientation from \ac{RSSI} and uses the predicted angle and its uncertainty to condition spatial processing, followed by temporal attention.} Table~\ref{tab:related_work_comparison} compares input requirements and validation settings.

\looseness=1\relax {Other studies apply learning to adaptive \ac{RIS} control. Chou et al.~\cite{chou2025drl} use deep reinforcement learning (DRL), specifically deep deterministic policy gradient (DDPG), to jointly optimize transmitter precoding and multiple \ac{RIS} configurations for multiuser downlink spectral efficiency. Their model includes reflection amplitude that depends on phase and a limited phase range. The policy uses information about user presence, previous actions, and rate feedback, and is evaluated through channel simulations. For discrete control, Wang et al.~\cite{wang2025heuristicdrl} combine a double deep Q-network (DDQN) with greedy refinement to optimize \ac{RIS} phase configurations through accumulated actions on individual columns. On our experimental platform, Kanaan et al.~\cite{kanaan2025multimodal} use camera observations and inertial coordinates to select \ac{RIS} and \ac{UE} beam codewords with DDQN and proximal policy optimization. These studies learn adaptive communication policies. The present work learns position and orientation from prescribed \ac{RSSI} sweeps.}

\subsection{Motivation and Contributions}

The preceding studies motivate an \ac{RSSI}-based estimator that infers orientation and accounts for its effect on fingerprints. {A practical application is positioning mobile robots or tracked assets in an indoor area served by a shared \ac{BS} and \ac{RIS}. Each device records a fingerprint under the common downlink illumination and reports its measurements to the \ac{BS}, which estimates position and orientation. Controlled offline mixtures inspired by power-domain \ac{NOMA}~\cite{wang2018noma} test the estimator when the power contributions of several \acp{UE} overlap in one observation. Section~\ref{sec:system_model} shows that downlink \ac{NOMA} with a common precoder and \ac{RIS} state leaves each device's mean fingerprint unchanged and defines the mixture construction.}

This paper develops the \ac{ODSG} framework to combine inferred orientation, spatial \ac{RSSI} structure, and temporal context. {The methodological contribution is the integration of \ac{FiLM}, \ac{MoE}, recurrent processing, and attention around a shared representation of inferred orientation. Each component builds on established methods~\cite{perez2018film,jacobs1991adaptive,graves2005framewise,vaswani2017attention} and serves a distinct role in localization.} The main contributions are as follows:

\begin{itemize}

\item \textbf{RSSI-derived orientation conditioning:} A lightweight \ac{CNN} estimates an orientation-class posterior from \ac{RSSI}. {The expected class angle and posterior entropy form the conditioning representation, replacing supplied orientation at inference while orientation labels supervise training.}

\item \textbf{Coupled spatial and temporal processing:} {Building on \ac{DSOG}~\cite{advani2025dsog}, the inferred orientation representation modulates spatial feature channels through \ac{FiLM} and informs soft expert gating together with the observed features. The spatial descriptors and a complementary statistical stream are processed by a \ac{BiLSTM} and temporal self-attention. This design separates feature adaptation, expert weighting, and temporal aggregation.}

\item \looseness=1\relax \textbf{Measurement-based evaluation with offline multi-user emulation:} We evaluate localization, orientation prediction, ablations, and inference cost on 28~GHz \ac{NLoS} measurements. {A separate offline experiment measures the accuracy and computational load of target-conditioned localization from power mixtures of single-user fingerprints as the number of emulated contributors grows, under the stated synthesis assumptions.}
\end{itemize}

\begin{figure}[!t]
\centering
\includegraphics[width=0.88\columnwidth]{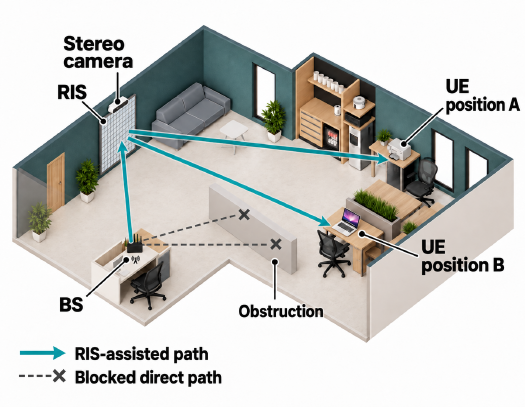}
\caption{RIS-assisted \ac{NLoS} localization. Reflected paths bypass the obstruction. {The illustrated UE positions represent separate acquisitions.}}
\label{fig:system_model}
\end{figure}

\subsection{Organization and Notation}

\looseness=1\relax Section~II presents the system model and problem formulation. Section~III details the \ac{ODSG} architecture. Section~IV presents the experimental setup, performance evaluation, and deployment discussion. Section~V concludes the paper. Table~\ref{tab:notation} summarizes the notation.

\begin{table}[!t]
\centering
\caption{Mathematical Notation}
\label{tab:notation}
\fontsize{8}{9.6}\selectfont
\setlength{\tabcolsep}{3pt}
\begin{tabularx}{\columnwidth}{@{}c>{\raggedright\arraybackslash}X@{}}
\toprule
\textbf{Symbol} & \textbf{Description} \\
\midrule
\multicolumn{2}{l}{\textit{General Conventions}} \\
$\mathbf{X}, \mathbf{x}, x$ & Matrix, vector, scalar \\
$(\cdot)^H, (\cdot)^T$ & Hermitian transpose, transpose \\
$\|\cdot\|$ & Euclidean norm \\
$|\cdot|$ & Absolute value/cardinality \\
$\odot$ & Hadamard (element-wise) product \\
$[\cdot \| \cdot]$ & Concatenation \\
$\text{diag}(\cdot)$ & Diagonal matrix \\
$\mathbb{E}[\cdot]$ & Expectation operator \\
$\mathsf{S}(\cdot)$ & Normalized Shannon entropy \\
$\sigma(\cdot), \delta(\cdot)$ & Sigmoid and ReLU activation functions \\
$\delta_H$ & Huber loss threshold parameter \\
$\mathbb{C}^{m \times n}, \mathbb{R}^{m \times n}$ & Complex/real $m \times n$ matrices \\
$\mathcal{CN}(\boldsymbol{\mu}, \boldsymbol{\Sigma})$ & Complex Gaussian distribution \\
$ (\cdot)^{(k)}, (\cdot)^{(n)}, (\cdot)_u$ & Snapshot index, training-sample index, target index \\
\midrule
\multicolumn{2}{l}{\textit{Key System Parameters}} \\
$N_B, N_U$ & Numbers of BS and UE antennas \\
$N_R$ & Number of RIS reflecting elements \\
$N_w, N_{\phi}$ & UE beams, RIS configurations \\
$T, U$ & Sequence length, number of users/contributors \\
$u$ & Target user/contributor index ($u \in \{1, \ldots, U\}$) \\
$\boldsymbol{\upsilon}_u^{(k)}, \varphi_u^{(k)}$ & UE position, orientation at {snapshot} $k$ \\
$\boldsymbol{\varepsilon}$ & Localization error vector \\
$\mathbf{R}_u^{(k)}$ & RSSI matrix for user $u$ at {snapshot} $k$ \\
$\mathbf{R}_{\mathrm{MU}}^{(k)}$ & Offline composite RSSI matrix \\
$\mathbf{X}_u^{(k)}$ & Network input for target index $u$ \\
$\boldsymbol{\Phi}_j, \mathbf{w}_i$ & RIS phase configuration, UE beam \\
$\eta_u^{(k)}, \gamma_{\mathrm{pl}}$ & Power fraction (mixture weight in (8)), path loss exponent \\
$h_{u,i,j}^{(k)}$ & Effective downlink channel of UE $u$ \\
$d_u$ & Distance from RIS to user $u$ \\
$\mathbf{e}_{\mathrm{user}}$ & Target-index embedding \\
$f_{\mathrm{ori}}, f_{\mathrm{loc}}$ & Orientation, localization networks \\
$\boldsymbol{\theta}_{\mathrm{ori}}, \boldsymbol{\theta}_{\mathrm{loc}}$ & Network parameters \\
$d_e, d_F$ & Embedding dimension, feature channels \\
$d_h, d_k$ & BiLSTM output, attention key dimensions \\
$S, k_s, \alpha_s$ & Number of scales, kernel sizes, fusion weights \\
$N_{s}$ & Number of training windows \\
$\mathcal{L}_{\mathrm{ori}}, \mathcal{L}_{\mathrm{loc}}$ & Orientation, localization losses \\
$\mathbf{W}, \mathbf{b}$ & Weight matrices and bias vectors \\
\bottomrule
\end{tabularx}
\end{table}

\section{System Model and Problem Formulation}
\label{sec:system_model}

\looseness=1\relax We consider a \ac{mmWave} localization system with a \ac{BS}, a \ac{RIS}, and \acp{UE} in an indoor \ac{NLoS} environment, as illustrated in Fig.~\ref{fig:system_model}. {The \acp{UE} measure downlink \ac{RSSI} during prescribed beam and \ac{RIS} sweeps and report their measurements to the \ac{BS}. The \ac{BS} performs position and orientation estimation. In the experiment, a single mobile \ac{UE} acquires a separate fingerprint at each measured location, and the multi-user extension combines these measurements offline.}
The \ac{BS} has a \ac{UPA} of $N_B$ antennas and a known, fixed position $\mathbf{p}_b \in \mathbb{R}^2$. The \ac{RIS} has $N_{R,v}$ rows and $N_{R,h}$ columns, comprising $N_R = N_{R,v}N_{R,h}$ passive reflecting elements, and a known position $\mathbf{p}_r \in \mathbb{R}^2$. Each \ac{UE} employs a \ac{UPA} with $N_U$ elements and is at an unknown Cartesian position $\boldsymbol{\upsilon}_u^{(k)} = [x_u^{(k)}, y_u^{(k)}]^\mathsf{T} \in \mathbb{R}^2$ with an unknown azimuthal orientation $\varphi_u^{(k)} \in [-\pi, \pi)$ at snapshot index $k$. {These coordinates denote horizontal projections onto the localization plane. From a window of $T$ snapshots, the framework outputs the endpoint position estimate $\hat{\boldsymbol{\upsilon}}_u^{(k+T-1)}$ and orientation estimate $\hat{\varphi}_u^{(k+T-1)}$.}

\subsection{Signal Model}\label{subsec:signal_model}

The \ac{BS}--\ac{RIS} channel, $\mathbf{H} \in \mathbb{C}^{N_R \times N_B}$, follows a geometric multipath model~\cite{shahmansoori2018posnavmmwave} with $L_H$ paths:
\begin{align}
    \mathbf{H} &= \sqrt{\frac{N_B N_R}{L_H}} \sum_{\ell=1}^{L_H} \alpha_H^{(\ell)} \nonumber\\
    &\quad\times \mathbf{a}_R(\phi_{H,R}^{(\ell)}, \theta_{H,R}^{(\ell)}) \mathbf{a}_B^H(\phi_{H,B}^{(\ell)}, \theta_{H,B}^{(\ell)}),
\end{align}
where $\alpha_H^{(\ell)} \in \mathbb{C}$ is the gain of path $\ell$. The unit-norm array responses $\mathbf{a}_B(\cdot)\in\mathbb{C}^{N_B}$ and $\mathbf{a}_R(\cdot)\in\mathbb{C}^{N_R}$ use azimuth/elevation pairs at the respective arrays: $(\phi_{H,B}^{(\ell)},\theta_{H,B}^{(\ell)})$ for departure from the \ac{BS} and $(\phi_{H,R}^{(\ell)},\theta_{H,R}^{(\ell)})$ for arrival at the \ac{RIS}. We approximate $\mathbf{H}$ as constant over an observation window, assuming fixed infrastructure and negligible environmental change.
The \ac{RIS}--\ac{UE} channel $\mathbf{G}_u^{(k)} \in \mathbb{C}^{N_U \times N_R}$ for user $u$ at snapshot $k$ contains $L_G$ multipath components and depends on the \ac{UE}'s position and orientation:
\begin{align}
    \mathbf{G}_u^{(k)} &= \sqrt{\frac{N_R N_U}{L_G}} \sum_{\ell=1}^{L_G} \alpha_{G,u}^{(\ell,k)} \nonumber\\
    &\quad\times \mathbf{a}_U(\phi_{G,U,u}^{(\ell,k)}, \theta_{G,U,u}^{(\ell,k)}) \mathbf{a}_R^{H}(\phi_{G,R,u}^{(\ell,k)}, \theta_{G,R,u}^{(\ell,k)}),
\end{align}
where $\alpha_{G,u}^{(\ell,k)} \in \mathbb{C}$ is the complex path gain and $\mathbf{a}_U(\cdot)\in\mathbb{C}^{N_U}$ is the unit-norm \ac{UE} array response. The subscripts $R$ and $U$ identify departure angles at the \ac{RIS} and arrival angles in the \ac{UE}'s local array coordinates, respectively. The arrival angles vary with device orientation. The direct \ac{BS}--\ac{UE} channel, $\mathbf{D}_u^{(k)} \in \mathbb{C}^{N_U \times N_B}$, is included for completeness and is considered negligible when an obstruction blocks the direct path.

For snapshot $k$, the \ac{BS} transmits a pilot symbol $s^{(k)} \in \mathbb{C}$ with average power $P_s$ using a unit-norm precoding vector $\mathbf{f} \in \mathbb{C}^{N_B}$. The fixed precoder steers toward the \ac{RIS} throughout the measurement sweep. The \ac{RIS} applies $\boldsymbol{\Phi}_j = \mathrm{diag}(e^{\mathrm{j}\phi_{j,1}}, \ldots, e^{\mathrm{j}\phi_{j,N_R}})$, where $\mathrm{j}^2=-1$ and $\phi_{j,n}$ is the phase shift of element $n$ in configuration $j$. This matrix is selected from a fixed codebook $\mathcal{C}_{\Phi} = \{\boldsymbol{\Phi}_1, \ldots, \boldsymbol{\Phi}_{N_{\phi}}\}$ that provides spatial diversity and is common to all users. {This nominal matrix represents phase shifts at unit reflection amplitude. The measured fingerprints retain the testbed's physical amplitude and phase response, including any coupling between them~\cite{abeywickrama2020practical,chou2025drl}.} The \ac{UE} uses a fixed receive codebook $\mathcal{C}_w = \{\mathbf{w}_1, \ldots, \mathbf{w}_{N_w}\}$ of unit-norm beams $\mathbf{w}_i \in \mathbb{C}^{N_U}$, also common to all users.
The baseband signal received at \ac{UE} $u$ with receive beam $\mathbf{w}_i$ and \ac{RIS} configuration $\boldsymbol{\Phi}_j$ {for snapshot} $k$ is
\begin{align}
    y_{u,i,j}^{(k)} = \mathbf{w}_i^H \Big(\mathbf{G}_u^{(k)} \boldsymbol{\Phi}_j \mathbf{H} + \mathbf{D}_u^{(k)}\Big) \mathbf{f} s^{(k)} + \mathbf{w}_i^H \mathbf{n}_{u,i,j}^{(k)},
\end{align}
where $\mathbf{n}_{u,i,j}^{(k)} \sim \mathcal{CN}(\mathbf{0}, \sigma^2 \mathbf{I}_{N_U})$ is \ac{AWGN} with variance $\sigma^2$ per receive element. {When several \acp{UE} are served in the considered power-domain \ac{NOMA} downlink, $s^{(k)}$ in~(3) is the superposition of their symbols, which share the precoder $\mathbf{f}$ and the \ac{RIS} state~\cite{wang2018noma}:}
\begin{equation}
     s^{(k)} =  \left[\sum_{u=1}^{U}\sqrt{\eta_u} {s}_u^{(k)}\right] \sqrt{P_s},
\end{equation}
where ${s}_u^{(k)}$ is a symbol for user $u$, with $\mathbb{E}[|s_u^{(k)}|^2]=1$, and $\{\eta_u\}_{u=1}^U$ are power allocation coefficients satisfying $\eta_u \ge 0$ and $\sum_u \eta_u = 1$.\footnote{{Power-allocation design, for example for rate or fairness, is outside the scope of this work. The localization framework neither estimates nor optimizes $\{\eta_u\}$ and operates with any downlink allocation policy.}}
The received power at \ac{UE} $u$ is
\begin{align}
    r_{u,i,j}^{(k)} = |y_{u,i,j}^{(k)}|^2. \label{eq:rx_power}
\end{align}
Assuming a quasi-static channel during the beam and \ac{RIS} sweep, the \ac{RSSI} matrix for snapshot $k$ is
\begin{align}
    \mathbf{R}_u^{(k)} = [r_{u,i,j}^{(k)}]_{i=1, \ldots, N_w}^{j=1, \ldots, N_{\phi}} \in \mathbb{R}_+^{N_w \times N_{\phi}}, \label{eq:rssi_matrix}
\end{align}
{where $\mathbf{R}_u^{(k)}$ is the fingerprint measured at \ac{UE} $u$ and reported to the \ac{BS}. Inter-user interference is part of the fingerprint but does not change its mean. All superimposed symbols reach \ac{UE} $u$ through the same effective channel $h_{u,i,j}^{(k)}=\mathbf{w}_i^H(\mathbf{G}_u^{(k)}\boldsymbol{\Phi}_j\mathbf{H}+\mathbf{D}_u^{(k)})\mathbf{f}$. For mutually uncorrelated zero-mean user symbols, the mean of~\eqref{eq:rx_power} over symbols and noise, for a given channel, is therefore}
\begin{align*}
{\mathbb{E}\big[r_{u,i,j}^{(k)}\big]}&{{}=P_s|h_{u,i,j}^{(k)}|^2\Big(\eta_u+\sum_{m\neq u}\eta_m\Big)+\sigma^2}\\
&{{}=P_s|h_{u,i,j}^{(k)}|^2+\sigma^2,}
\end{align*}
{where $P_s|h_{u,i,j}^{(k)}|^2\sum_{m\neq u}\eta_m$ is the inter-user interference power. The power split affects decoding, \ac{SINR}, rate, and bit error rate, whereas the mean fingerprint equals that of a single-user transmission or a common pilot at the same total power, for any number of users and any allocation. Power averaging at a linear receiver suppresses the cross terms between mutually uncorrelated waveforms~\cite{proakis2008digital}, and finite averaging intervals can retain fluctuations around the mean. Consequently, single-user measurements at the corresponding positions and orientations represent the mean fingerprints of simultaneously served \acp{UE}, provided that the other devices do not alter the propagation channels. The equality requires all streams to share the precoder and \ac{RIS} state. With user-specific precoders, each fingerprint would depend on the other users' precoders and power fractions.}

{One possible allocation policy assigns larger power fractions to more distant \acp{UE} to compensate path loss. The offline evaluation uses the normalized distance-based rule below to set the relative powers of the mixture contributions:}
\begin{align}
    {\eta_u^{(k)} = \frac{(d_u^{(k)})^{\gamma_{\mathrm{pl}}}}{\sum_{m=1}^U (d_m^{(k)})^{\gamma_{\mathrm{pl}}}},} \label{eq:distance_weights}
\end{align}
where $d_u^{(k)} = \|\boldsymbol{\upsilon}_u^{(k)} - \mathbf{p}_r\|$ is the horizontal distance from the \ac{RIS} to user $u$ and $\gamma_{\mathrm{pl}}$ is the assumed path loss exponent. {The weights are recomputed at each snapshot from the ground-truth positions available for offline synthesis. For positive distances, the ratio $\eta_a^{(k)}/\eta_b^{(k)}=(d_a^{(k)}/d_b^{(k)})^{\gamma_{\mathrm{pl}}}$ sets a reproducible power imbalance for the evaluation.}

{The offline extension combines independently measured matrices in linear power units to form the common composite}
\begin{equation}
{\mathbf{R}_{\mathrm{MU}}^{(t)}=\sum_{m=1}^{U}\eta_m^{(t)}\mathbf{R}_m^{(t)}.}
\label{eq:offline_mixture}
\end{equation}
{The composite $\mathbf{R}_{\mathrm{MU}}^{(t)}\in\mathbb{R}_+^{N_w\times N_\phi}$ is shared by the emulated targets, with weights from~\eqref{eq:distance_weights}. Its noise contribution comes entirely from the measured matrices and thus reflects the separate acquisitions. Normalized weights preserve a common mean noise floor. The term \ac{NOMA}-style refers specifically to this offline construction.}

\subsubsection{Deployment Motivation and Scope}
\label{subsec:deployment_scope}
{Localization from the reported downlink fingerprints is decoupled from power allocation. Because the mean fingerprint at each \ac{UE} does not depend on the power fractions, the \ac{BS} processes each report as a single-user input under any allocation policy, without requiring the current positions. If the \ac{BS} applies the distance-based rule in~\eqref{eq:distance_weights}, it can use the most recent estimate $\hat{\boldsymbol{\upsilon}}_{u,\mathrm{prev}}$ of each \ac{UE} $u$. Substituting $\hat d_{u,\mathrm{prev}}=\|\hat{\boldsymbol{\upsilon}}_{u,\mathrm{prev}}-\mathbf p_r\|$ into~\eqref{eq:distance_weights} gives normalized fractions for subsequent downlink transmissions. A preset normalized allocation applies until the first estimate is available. New \ac{RSSI} reports then update the position and orientation estimates at the \ac{BS}.}

\looseness=1\relax {Unlike the per-\ac{UE} fingerprint in~\eqref{eq:rssi_matrix}, the offline sum in~\eqref{eq:offline_mixture} superimposes fingerprints measured at distinct \ac{UE} locations in one observation. In mean power, the sum matches the observation at a common receiver of mutually uncorrelated simultaneous transmissions from several \acp{UE} through distinct channels, for example the \ac{BS} in power-domain uplink \ac{NOMA} with reciprocal channels, \ac{UE} beams swept in step with the \ac{RIS} configurations, and power shares $\eta_m$. The other contributions then change the observed \ac{RSSI}, which measures the total received power, and the estimator must localize each requested contributor from the shared composite. The common receiver acquires the composite for all $U$ contributors in one sweep. Concurrent measurements at such a receiver are needed to quantify the effects of synchronization offsets, simultaneous blockage, receiver noise, saturation, gain control, and channel access on localization accuracy and acquisition latency.}

\begin{figure*}[!t]
  \centering
  \includegraphics[width=0.8\textwidth]{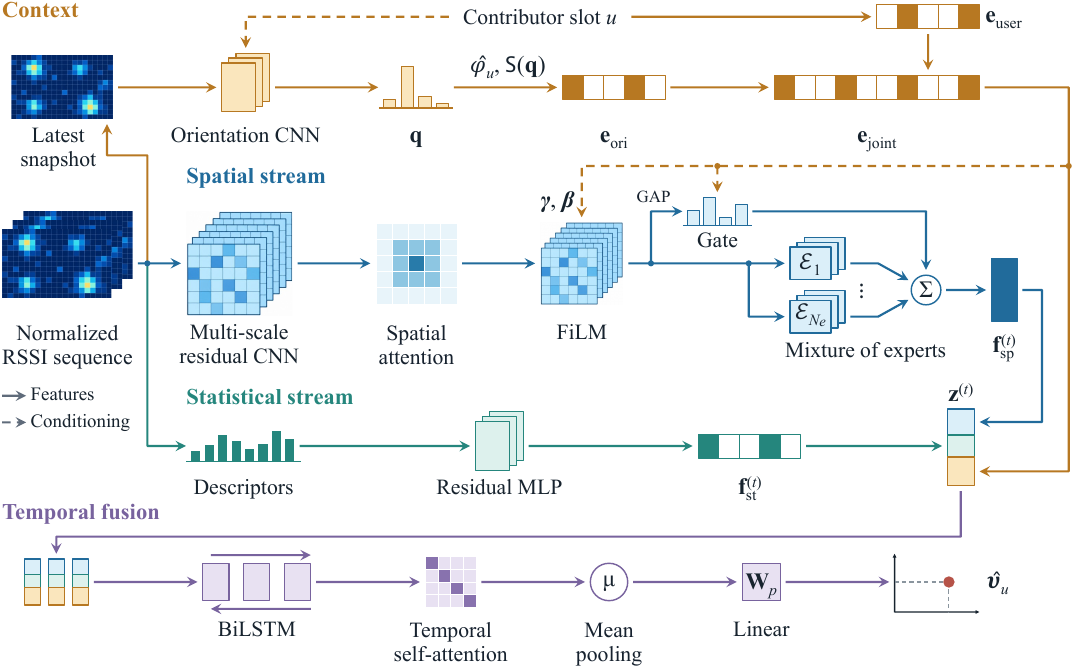}
  \caption{Architecture of the \ac{ODSG} framework, integrating orientation prediction, dual-stream feature extraction, and temporal processing for indoor localization. {The target user index denotes the contributor slot defined in Section~II.} Statistical descriptors summarize normalized grid entries.}
  \label{fig:model_architecture}
\end{figure*}

\subsection{Learning Tasks and Problem Formulation}

The central objective is to estimate the target \ac{UE}'s final position, $\boldsymbol{\upsilon}_u^{(k+T-1)}$, and orientation, $\varphi_u^{(k+T-1)}$, from a sequence of $T$ input matrices. {The network input $\mathbf{X}_u^{(t)}$ is the measured $\mathbf{R}_u^{(t)}$ in single-user evaluation and the common composite $\mathbf{R}_{\mathrm{MU}}^{(t)}$ in offline multi-user evaluation. The subscript $u$ identifies the requested target.}
The training tuples are $\big(\{\mathbf{X}_u^{(t)}\}_{t=k}^{k+T-1},u,\varphi_u^{(k+T-1)},\boldsymbol{\upsilon}_u^{(k+T-1)}\big)$. {Contributors are indexed by increasing endpoint distance $d_u^{(k+T-1)}$, with ties resolved by endpoint $x$ and then $y$. This ordering is fixed throughout each window and used in all data partitions. Thus, $u$ denotes a contributor's rank within a mixture. Endpoint ordering provides a consistent label convention aligned with the orientation network's input from the final snapshot. Unique recovery of individual contributors depends on the information available in the mixture. Associating these contributors with physical devices requires an additional identification mechanism.}

\looseness=1\relax {Let $\widetilde{\mathbf{X}}_u^{(t)}$ denote the grid after normalization using training statistics, as defined in Section~\ref{sec:dsog}.} An orientation network, $f_{\mathrm{ori}}(\cdot; \boldsymbol{\theta}_{\mathrm{ori}})$, takes the final normalized \ac{RSSI} matrix and target index as inputs. It outputs a probability distribution, $\mathbf{q}_u^{(k+T-1)}$, over orientation classes indexed by $\Omega$, with representative angles $\varphi_c$. The orientation estimate is the posterior mean:
\begin{align}
    \mathbf{q}_u^{(k+T-1)} &= f_{\mathrm{ori}}(\widetilde{\mathbf{X}}_u^{(k+T-1)}, u; \boldsymbol{\theta}_{\mathrm{ori}}), \label{eq:ori_prob} \\
    \hat{\varphi}_u^{(k+T-1)} &= \sum_{c \in \Omega} \varphi_c \, q_{u,c}^{(k+T-1)},
\end{align}
\looseness=1\relax where $q_{u,c}^{(k+T-1)}$ is the predicted probability of class angle $\varphi_c$, and $\boldsymbol{\theta}_{\mathrm{ori}}$ denotes the learnable parameters of the orientation network. {The expected class angle is both the scalar orientation estimate and an input to the conditioning stage.} A localization network, $f_{\mathrm{loc}}(\cdot; \boldsymbol{\theta}_{\mathrm{loc}})$, then maps the full normalized \ac{RSSI} sequence, the orientation estimate, posterior entropy, and the target index to the final position
\begin{align}
    \hat{\boldsymbol{\upsilon}}_u^{(k+T-1)} = f_{\mathrm{loc}}\Big(&\{\widetilde{\mathbf{X}}_u^{(t)}\}_{t=k}^{k+T-1}, \hat{\varphi}_u^{(k+T-1)}, \nonumber\\
    &\mathsf{S}(\mathbf{q}_u^{(k+T-1)}), u; \boldsymbol{\theta}_{\mathrm{loc}}\Big).
\end{align}
{Here $\mathsf{S}(\mathbf{q})=-\sum_{c\in\Omega}q_c\log q_c/\log|\Omega|$ is normalized posterior entropy.}

{Position and orientation are outputs of the framework. At inference, the \ac{BS} processes the \ac{RSSI} sequence and target index, with orientation inferred internally. The fixed sweep operates independently of the unknown position. Ground-truth positions and orientations provide training labels, and the positions also determine the offline mixture weights.}
{Let $N_s$ denote the number of training windows and $U_n$ the number of contributors in window $n$, with $U_n=1$ for a measured single-user window. Each window contributes the mean loss over its targets, giving all windows equal weight regardless of the number of contributors.} Orientation estimation uses the class-weighted cross-entropy loss
\begin{align}
    {\mathcal{L}_{\mathrm{ori}} = - \frac{1}{N_s} \sum_{n=1}^{N_s}\frac{1}{U_n}\sum_{u=1}^{U_n}\sum_{c\in\Omega}
    w_c\mathbf{1}\{c=c_{n,u}\}\log q_{u,c}^{(n)},}
\end{align}
where $c_{n,u}$ is the ground-truth endpoint class of target $u$ in window $n$, $q_{u,c}^{(n)}$ is its predicted class probability, $w_c$ is the inverse training frequency weight for class $c$, and $\mathbf{1}\{\cdot\}$ is the indicator function. Position regression uses the Huber loss to reduce sensitivity to large errors:
\begin{align}
    {\mathcal{L}_{\mathrm{loc}} = \frac{1}{N_s} \sum_{n=1}^{N_s}\frac{1}{U_n}\sum_{u=1}^{U_n}
    \mathcal{L}_{\mathrm{Huber}}\big(\hat{\boldsymbol{\upsilon}}_u^{(n)} - \boldsymbol{\upsilon}_u^{(n)}; \delta_H\big),}
\end{align}
where the Huber loss for an error vector $\boldsymbol{\varepsilon} = [\varepsilon_x, \varepsilon_y]^\T$ is
\begin{align}
    \mathcal{L}_{\mathrm{Huber}}(\boldsymbol{\varepsilon}; \delta_H) = \sum_{d \in \{x,y\}} \begin{cases} \frac{1}{2} \varepsilon_d^2, & |\varepsilon_d| \leq \delta_H, \\ \delta_H (|\varepsilon_d| - \frac{1}{2} \delta_H), & |\varepsilon_d| > \delta_H. \end{cases}
\end{align}
The positive threshold $\delta_H$ separates the quadratic and linear regimes. The total loss is
\begin{align}
    \mathcal{L}_{\mathrm{total}} = \lambda_{\mathrm{ori}} \mathcal{L}_{\mathrm{ori}} + \lambda_{\mathrm{loc}} \mathcal{L}_{\mathrm{loc}},
\end{align}
with weighting factors $\lambda_{\mathrm{ori}}, \lambda_{\mathrm{loc}} \in [0,1]$ satisfying $\lambda_{\mathrm{ori}} + \lambda_{\mathrm{loc}} = 1$. After pretraining the orientation module, both networks are jointly fine-tuned by minimizing the total loss:
\begin{align}
    \min_{\boldsymbol{\theta}_{\mathrm{ori}}, \boldsymbol{\theta}_{\mathrm{loc}}} \mathcal{L}_{\mathrm{total}}. \label{OP}
\end{align}

\section{Proposed \acs{ODSG} Deep Learning Framework}
\label{sec:dsog}

The \ac{ODSG} framework in Fig.~\ref{fig:model_architecture} maps sequences of beam--\ac{RIS} power grids to position coordinates. {Its design extends the dual-stream recurrent architecture with orientation gating in~\cite{advani2025dsog}.} The expected angle and posterior entropy summarize the predicted orientation distribution. This representation conditions feature modulation and soft expert weighting, with the latter also informed by the spatial features. Temporal self-attention then aggregates the recurrent features over the observation window. {\ac{FiLM} adapts features to orientation, the expert gate combines distinct spatial representations, and temporal attention uses context across snapshots.}

\subsection{Input Representation and Normalization}
\looseness=1\relax Let $\mathbf{X}_u^{(t)} \in \mathbb{R}_+^{N_w \times N_\phi}$ denote the network input for target $u$ (Section~\ref{sec:system_model}), indexed by the receive beams and \ac{RIS} configurations of the measured fingerprints. The input sequence is $\mathcal{X}_u = \{\mathbf{X}_u^{(t)}\}_{t=k}^{k+T-1}$, with powers in linear units (watts) before preprocessing. {Both feature streams and the orientation branch use the normalized grid $\widetilde{\mathbf{X}}_u^{(t)}=(\mathbf{X}_u^{(t)}-\mu_{\mathrm{tr}})/\sigma_{\mathrm{tr}}$. The scalar mean $\mu_{\mathrm{tr}}$ and standard deviation $\sigma_{\mathrm{tr}}$ are fitted to the training measurements before forming overlapping windows and then fixed.}
Each \ac{RSSI} snapshot is obtained by sweeping over all $N_w$ receive beams and $N_\phi$ \ac{RIS} configurations. {A temporal window contains $T$ such snapshots, so a fresh window requires $TN_wN_\phi$ pilot observations. When the window advances by one snapshot per sweep, each later update reuses the preceding $T-1$ snapshots and requires $N_wN_\phi$ new observations. With an observation duration $T_p$ per beam--configuration pair, one sweep contains $N_wN_\phi T_p$ of observation time. Equations~\eqref{eq:sweep_latency} and~\eqref{eq:first_latency} relate these counts to acquisition time, reporting, and inference.}

\subsection{Integrated Orientation Prediction Module}
In \ac{mmWave} systems with planar arrays, the \ac{UE} orientation $\varphi_u$ alters the observed spatial power pattern. A dedicated orientation branch captures this dependence. Its lightweight \ac{CNN}, $f_{\mathrm{ori}}(\cdot;\boldsymbol{\theta}_{\mathrm{ori}})$, processes the most recent normalized snapshot $\widetilde{\mathbf{X}}_u^{(k+T-1)}$ and the target index $u$. It produces the posterior $\mathbf{q}_u^{(k+T-1)} \in [0,1]^{|\Omega|}$ over the orientation classes $\Omega$:
\begin{align}
    \mathbf{q}_u^{(k+T-1)} = f_{\mathrm{ori}}(\widetilde{\mathbf{X}}_u^{(k+T-1)}, u; \boldsymbol{\theta}_{\mathrm{ori}}).
\end{align}
As in~\eqref{eq:ori_prob}, $f_{\mathrm{ori}}$ includes its output Softmax. An orientation embedding $\mathbf{e}_{\mathrm{ori}} \in \mathbb{R}^{d_e}$ combines the predicted angle with posterior entropy as orientation and uncertainty context:
\begin{align}
    \mathbf{e}_{\mathrm{ori}} = \mathbf{W}_e
    \begin{bmatrix}
        \cos\hat{\varphi}_u^{(k+T-1)} \\
        \sin\hat{\varphi}_u^{(k+T-1)} \\
        \mathsf{S}(\mathbf{q}_u^{(k+T-1)})
    \end{bmatrix}{+\mathbf{b}_e},
\end{align}
where $\hat{\varphi}_u^{(k+T-1)}$ is the expected class angle, $\mathsf{S}(\mathbf{q})$ is the normalized entropy, $d_e$ is the embedding dimension, and $\mathbf{W}_e \in \mathbb{R}^{d_e \times 3}$ is a learnable projection matrix {with bias $\mathbf{b}_e\in\mathbb{R}^{d_e}$}.

{As shown in Fig.~\ref{fig:model_architecture}, the latest snapshot supplies one orientation context per window. For clarity, we omit its endpoint index from $\mathbf{e}_{\mathrm{ori}}$ and $\mathbf{e}_{\mathrm{joint}}$. This context is reused at each time step, so the \ac{FiLM} coefficients $\boldsymbol{\gamma}$ and $\boldsymbol{\beta}$ are fixed within the window. Expert weights can vary with $t$ because the gate also receives pooled features from each snapshot.}

\subsubsection{Conditioning for Emulated Multi-User Scenarios}
For the offline mixtures, a learnable embedding maps the target contributor index $u$ to $\mathbf{e}_{\mathrm{user}} \in \mathbb{R}^{d_e}$. Concatenating this vector with the orientation embedding gives $\mathbf{e}_{\mathrm{joint}} = [\mathbf{e}_{\mathrm{ori}} \| \mathbf{e}_{\mathrm{user}}] \in \mathbb{R}^{2d_e}$. {This vector conditions the estimate on the requested contributor under the emulation's labeling convention. Single-user processing uses the same embedding lookup with a fixed contributor slot, preserving the dimension of the joint context.}

{The index encodes a contributor's rank rather than a device identity. A new device therefore uses an existing index, namely the fixed single-user slot for its own reports or its rank within a mixture. The trained indices cover mixtures of up to eight contributors, and extending the table to a larger mixture requires learning the corresponding embeddings. Deployment with new physical devices requires consistent index association.}

\begin{figure*}[!t]
\centering
\includegraphics[width=0.75\textwidth]{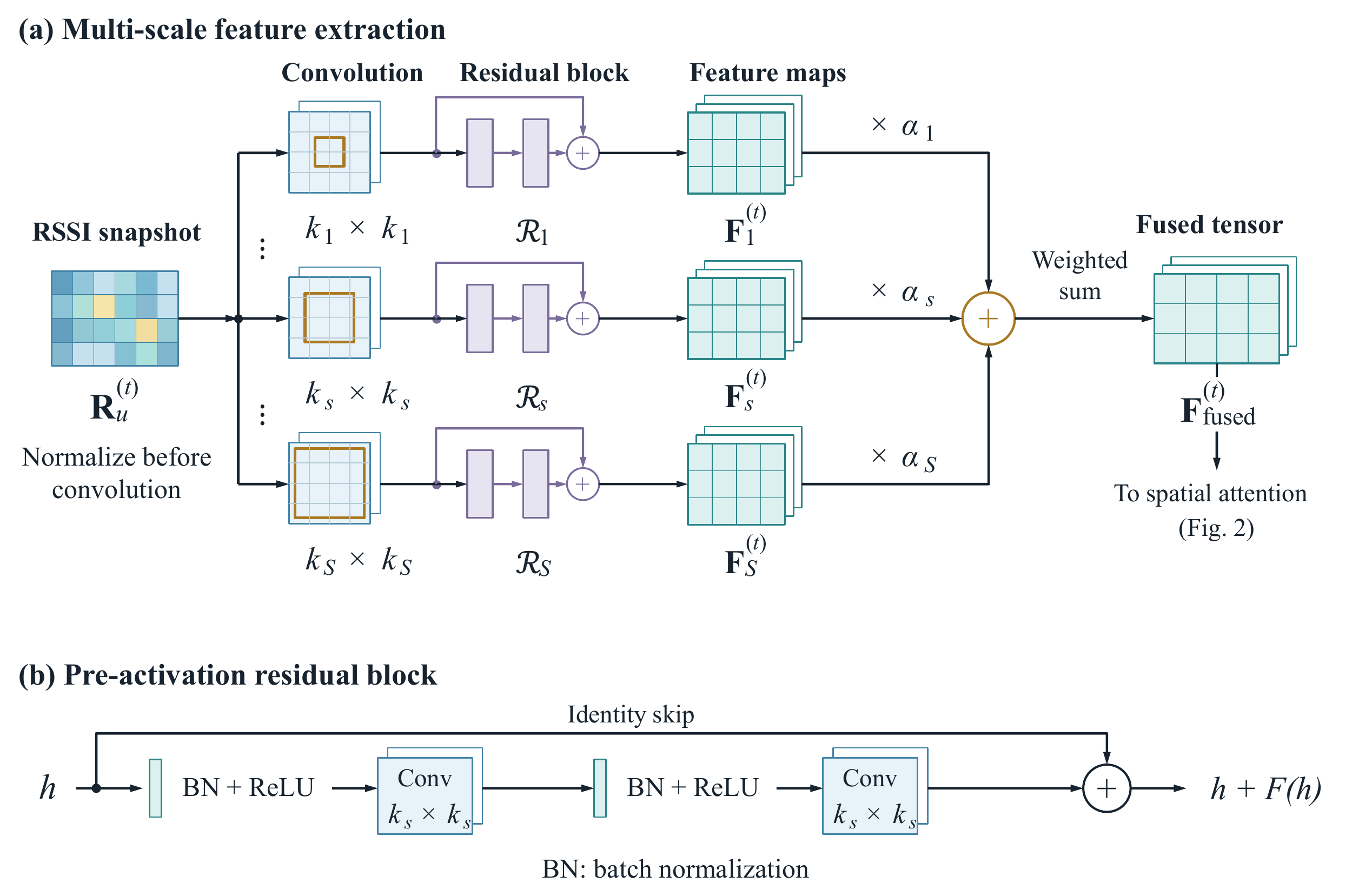}
\caption{Multi-scale residual \ac{CNN} architecture for spatial feature extraction from \ac{RSSI} matrices. Panel (a) shows weighted fusion. Its $\mathbf{R}_u^{(t)}$ label denotes the measured single-user input, corresponding to $\mathbf{X}_u^{(t)}$ before preprocessing. Normalization precedes the illustrated spatial operators, which also process the composite input in emulation. In panel (b), the local symbols $h$ and $F(h)$ denote the block input tensor and its learned residual transformation.}
\label{fig:residual_cnn}
\end{figure*}

\subsection{Spatial Feature Extraction Network}
The spatial branch extracts hierarchical beam--\ac{RIS} patterns.

\subsubsection{Multi-Scale Residual Processing}
The \ac{RSSI} matrix contains features at multiple scales, including sharp peaks from specular paths and diffuse clusters from scattering. To capture them, we use a multi-scale architecture with $S$ parallel convolutional branches having kernel sizes $k_s$. Each branch $s$ maps the normalized input $\widetilde{\mathbf{X}}_u^{(t)}$ to
\begin{align}
    \mathbf{F}_s^{(t)} = \mathcal{R}_s(\mathcal{C}_{k_s}(\widetilde{\mathbf{X}}_u^{(t)})),
\end{align}
\looseness=1\relax where $\mathcal{C}_{k_s}(\cdot)$ denotes convolution with a $k_s\times k_s$ kernel. The mapping $\mathcal{R}_s(\cdot)$ is a pre-activation residual block~\cite{he2016identity} with batch normalization (BN) and \ac{ReLU} activation $\delta(\cdot)$. Residual connections follow~\cite{he2016deep}. Branch outputs retain the receive-beam and \ac{RIS}-configuration dimensions and contain $d_F$ feature channels. Learnable scalar weights $\{\alpha_s\}$ fuse these outputs (Fig.~\ref{fig:residual_cnn}):
\begin{align}
    \mathbf{F}_{\text{fused}}^{(t)} = \sum_{s=1}^{S} \alpha_s \mathbf{F}_s^{(t)}.
\end{align}

\subsubsection{Spatial Attention Mechanism}
To mitigate the impact of noise and deep fading, a spatial attention module computes a 2D weighting map $\mathbf{A}_{\text{spatial}}^{(t)} \in [0,1]^{N_w \times N_\phi}$ that highlights informative regions of the feature grid. The map uses pointwise convolutions parameterized by weights $\mathbf{W}_{a1}, \mathbf{W}_{a2}$ and biases $\mathbf{b}_{a1}, \mathbf{b}_{a2}$:
\begin{align}
    \mathbf{A}_{\text{spatial}}^{(t)} = \sigma\left(\mathbf{W}_{a2} \delta(\mathbf{W}_{a1} \mathbf{F}_{\text{fused}}^{(t)} + \mathbf{b}_{a1}) + \mathbf{b}_{a2}\right),
\end{align}
where $\sigma(\cdot)$ is the Sigmoid function. The pointwise maps act along the feature channels and produce one weight per grid position, which is broadcast across channels:
\begin{align}
    \mathbf{F}_{\text{attended}}^{(t)} = \mathbf{F}_{\text{fused}}^{(t)} \odot \mathbf{A}_{\text{spatial}}^{(t)}.
\end{align}

\subsection{Feature-wise Linear Modulation and Mixture of Experts}
{\Ac{FiLM} applies a learned affine transformation to each channel of the attended features, $\mathbf{F}_{\mathrm{attended}}^{(t)}$~\cite{perez2018film}.} The joint embedding conditions this transformation before expert processing:
\begin{align}
  \boldsymbol{\gamma} &= \mathbf{W}_\gamma \mathbf{e}_{\mathrm{joint}} + \mathbf{b}_\gamma, \\
  \boldsymbol{\beta} &= \mathbf{W}_\beta \mathbf{e}_{\mathrm{joint}} + \mathbf{b}_\beta, \\
  \mathbf{F}_{\mathrm{mod}}^{(t)} &= \boldsymbol{\gamma} \odot \mathbf{F}_{\mathrm{attended}}^{(t)} + \boldsymbol{\beta}.
\end{align}
The learnable matrices are $\mathbf{W}_\gamma,\mathbf{W}_\beta\in\mathbb{R}^{d_F\times 2d_e}$, and the biases are $\mathbf{b}_\gamma,\mathbf{b}_\beta\in\mathbb{R}^{d_F}$. The affine transformation broadcasts across the spatial grid.
{A \ac{MoE} subsequently combines specialized feature transformations using soft gating~\cite{jacobs1991adaptive}. Whereas \ac{FiLM} rescales and shifts feature channels, the gate weights the outputs of distinct expert networks.} The weights $\mathbf{g}^{(t)}$ for $N_e$ experts, $\{\mathcal{E}_j\}_{j=1}^{N_e}$, depend on both the joint embedding and a pooled feature summary:
\begin{align}
  \mathbf{g}^{(t)} &= \mathrm{Softmax}\Big(\mathbf{W}_{g2} \nonumber\\
  &\quad\delta\big(\mathbf{W}_{g1} [\mathbf{e}_{\mathrm{joint}} \,\|\, \bar{\mathbf{f}}^{(t)}] + \mathbf{b}_{g1}\big) + \mathbf{b}_{g2}\Big), \\
  \mathbf{f}_j^{(t)} &= \mathcal{E}_j(\mathbf{F}_{\mathrm{mod}}^{(t)}) \in \mathbb{R}^{d_{sp}}, \quad j = 1, \ldots, N_e, \\
  \mathbf{f}_{\mathrm{sp}}^{(t)} &= \sum_{j=1}^{N_e} g_j^{(t)} \mathbf{f}_j^{(t)},
\end{align}
where $\bar{\mathbf{f}}^{(t)} = \mathrm{GlobalAvgPool}(\mathbf{F}_{\mathrm{mod}}^{(t)}) \in \mathbb{R}^{d_F}$ averages each channel over the grid. The learnable matrices are $\mathbf{W}_{g1} \in \mathbb{R}^{d_g \times (2d_e + d_F)}$ and $\mathbf{W}_{g2} \in \mathbb{R}^{N_e \times d_g}$, with hidden dimension $d_g$ and biases $\mathbf{b}_{g1} \in \mathbb{R}^{d_g}$ and $\mathbf{b}_{g2} \in \mathbb{R}^{N_e}$. Each expert output has dimension $d_{sp}$, and the Softmax gives $g_j^{(t)}>0$ and $\sum_j g_j^{(t)}=1$, so every expert contributes. {Conditioning on orientation and observed features allows different expert combinations for similar estimated orientations.}

\subsection{Statistical Aggregation Branch}
In parallel with spatial processing, a statistical branch summarizes the {normalized power grid through dimensionless descriptors.} Its first descriptor sums all grid entries:
\begin{align}
    s_{\mathrm{tot}}^{(t)} = \sum_{i,j} \widetilde{\mathbf{X}}_u^{(t)}(i,j).
\end{align}
The dynamic range captures the spread of normalized values:
\begin{align}
    \Delta_{\mathrm{dyn}}^{(t)} = \max_{i,j} \widetilde{\mathbf{X}}_u^{(t)}(i,j) - \min_{i,j} \widetilde{\mathbf{X}}_u^{(t)}(i,j).
\end{align}
Summing across \ac{RIS} configurations gives $\mathbf{s}_w^{(t)}\in\mathbb{R}^{N_w}$:
\begin{align}
    \mathbf{s}_w^{(t)}(i) = \sum_j \widetilde{\mathbf{X}}_u^{(t)}(i,j), \quad i = 1, \ldots, N_w.
\end{align}
Summing across beams gives $\mathbf{s}_{\phi}^{(t)}\in\mathbb{R}^{N_\phi}$:
\begin{align}
    \mathbf{s}_{\phi}^{(t)}(j) = \sum_i \widetilde{\mathbf{X}}_u^{(t)}(i,j), \quad j = 1, \ldots, N_{\phi}.
\end{align}
A multilayer perceptron $\mathcal{H}_{\mathrm{st}}$ with residual connections maps these $N_w+N_\phi+2$ statistics to $\mathbf{f}_{\mathrm{st}}^{(t)}$ of dimension $d_{st}$:
\begin{align}
  \mathbf{f}_{\mathrm{st}}^{(t)} = \mathcal{H}_{\mathrm{st}}([s_{\mathrm{tot}}^{(t)} \,\|\, \mathbf{s}_w^{(t)} \,\|\, \mathbf{s}_{\phi}^{(t)} \,\|\, \Delta_{\mathrm{dyn}}^{(t)}]) \in \mathbb{R}^{d_{st}}.
\end{align}

\subsection{Temporal Fusion and Localization Head}
We concatenate the spatial features, statistical features, and joint orientation and contributor context at each time step $t$:
\begin{align}
    \mathbf{z}^{(t)} = \left[ \mathbf{f}_{\mathrm{sp}}^{(t)} \,\|\, \mathbf{f}_{\mathrm{st}}^{(t)} \,\|\, \mathbf{e}_{\mathrm{joint}} \right] \in \mathbb{R}^{d_{sp} + d_{st} + 2d_e}.
\end{align}
A \ac{BiLSTM}~\cite{graves2005framewise} processes the sequence $\{\mathbf{z}^{(t)}\}_{t=k}^{k+T-1}$. Its forward and backward passes remain within the acquired window. Stacking the outputs as rows gives
\begin{align}
   \mathbf{B} = \mathrm{BiLSTM}(\mathbf{z}^{(k)}, \ldots, \mathbf{z}^{(k+T-1)})\in\mathbb{R}^{T\times d_h},
\end{align}
where $d_h$ is the output dimension. {A stack of $L_A$ temporal attention layers processes these features. Each layer combines multi-head self-attention~\cite{vaswani2017attention}, a positionwise feedforward network with hidden width $d_{\mathrm{ff}}$, and layer normalization. For a layer input $\mathbf{C}\in\mathbb{R}^{T\times d_h}$, head $i$ computes}
\begin{align}
    {\mathbf{Head}_i(\mathbf{C})} &= {\mathrm{Softmax}\left(\frac{\mathbf{C} \mathbf{W}_Q^i (\mathbf{C} \mathbf{W}_K^i)^T}{\sqrt{d_k}}\right) \mathbf{C} \mathbf{W}_V^i,}
\end{align}
where Softmax acts along each row of the $T\times T$ score matrix. The $N_h$ heads are concatenated and projected:
\begin{align}
    {\mathrm{MHA}(\mathbf{C})} &= {[\mathbf{Head}_1(\mathbf{C}) \| \dots \| \mathbf{Head}_{N_h}(\mathbf{C})] \mathbf{W}_O,}
\end{align}
\begin{figure*}[!t]
\centering
\includegraphics[width=0.8\textwidth]{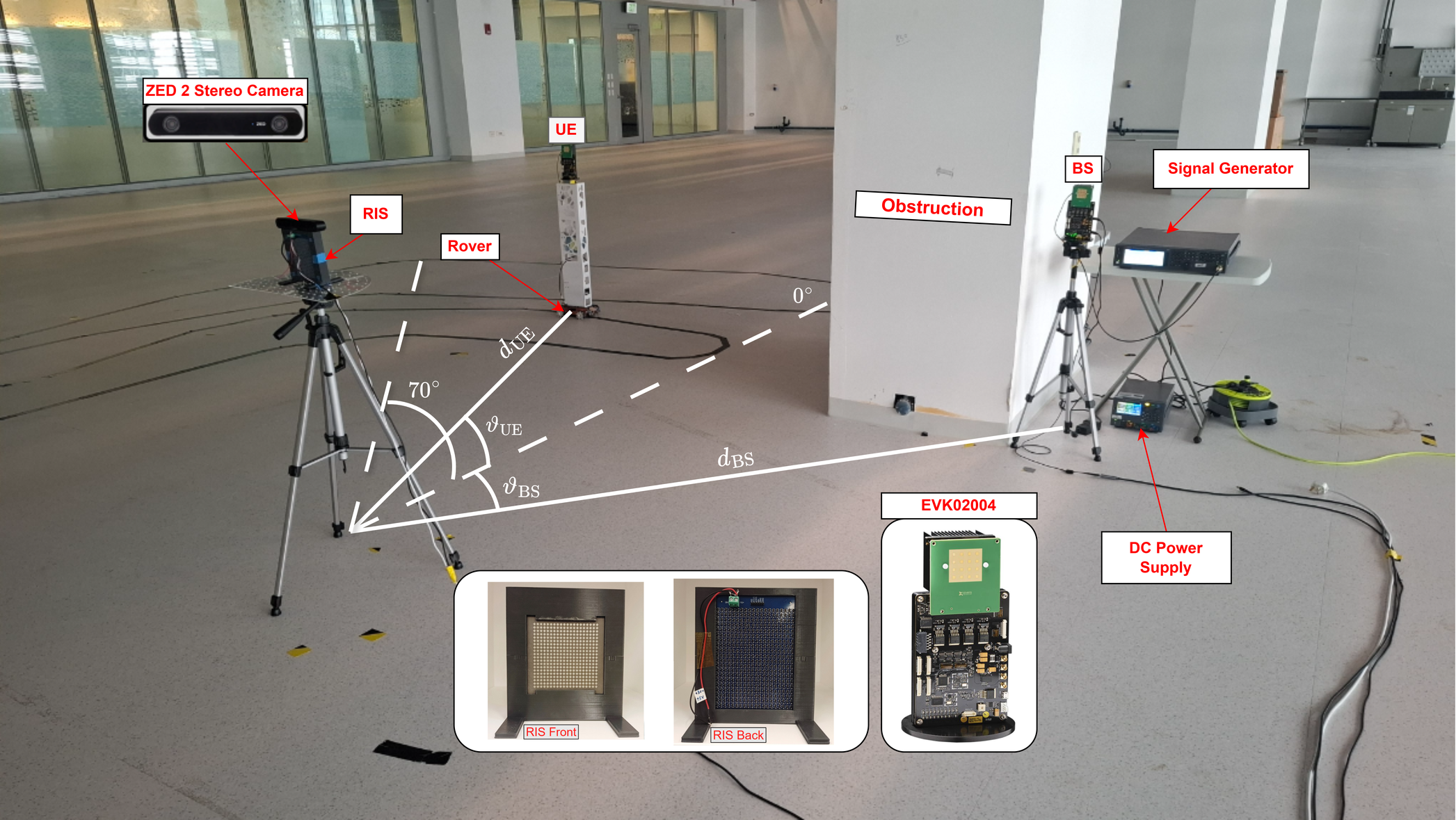}
\caption{The experimental testbed for \ac{RIS}-aided \ac{mmWave} localization.}
\label{fig:experimental_setup}
\end{figure*}%
\looseness=1\relax where $\mathbf{W}_Q^i, \mathbf{W}_K^i, \mathbf{W}_V^i \in \mathbb{R}^{d_h \times d_k}$ are learned query, key, and value projections with common dimension $d_k$. The output projection $\mathbf{W}_O \in \mathbb{R}^{N_h d_k \times d_h}$ preserves the input dimensions. {Projection biases are omitted for brevity. Composing the independently parameterized layers $\mathcal{A}_{\ell}$ gives $\mathbf{B}_{\mathrm{att}}=(\mathcal{A}_{L_A}\circ\cdots\circ\mathcal{A}_1)(\mathbf{B})\in\mathbb{R}^{T\times d_h}$.} Temporal mean pooling gives
\begin{align}
    \bar{\mathbf{h}} &= \mathrm{MeanPool}(\mathbf{B}_{\mathrm{att}}),
\end{align}
where $\mathrm{MeanPool}(\cdot)$ averages the rows into the column vector $\bar{\mathbf{h}}\in\mathbb{R}^{d_h}$. A linear head predicts the endpoint position:
\begin{align}
    \hat{\boldsymbol{\upsilon}}_u^{(k+T-1)} = \mathbf{W}_p \bar{\mathbf{h}} + \mathbf{b}_p,
\end{align}
where $\mathbf{W}_p \in \mathbb{R}^{2 \times d_h}$ is the output projection matrix for the 2D position coordinates and $\mathbf{b}_p \in \mathbb{R}^2$ is the bias vector.

\subsection{Multi-Task Loss and Training}
After orientation pretraining, the two networks are jointly optimized using the composite loss defined in Section~\ref{sec:system_model}. {Training starts with single-user inputs and then adds offline mixtures with increasing numbers of contributors.}

\subsection{Computational Complexity}
\label{subsec:complexity}
{Let $C_{\mathrm{fr}}$ denote the operation count for spatial extraction, statistical aggregation, and conditioning at one time step, excluding orientation inference. A convolutional layer $\ell$ contributes proportionally to $A_\ell k_\ell^2 c_{\ell,\mathrm{in}}c_{\ell,\mathrm{out}}$, where $A_\ell$ is its output grid area, $k_\ell$ is its kernel side length, and $c_{\ell,\mathrm{in}}$ and $c_{\ell,\mathrm{out}}$ are its input and output channel counts. Summing over all convolutional layers includes the parallel scales and all $N_e$ experts, because the soft \ac{MoE} evaluates every expert.}

{For a batch of $M$ sequences, each conditioned on its target index, the operation count is bounded by}
\begin{align}
&{\mathcal{O}\!\Big(M\big[ C_{\mathrm{ori}}+T C_{\mathrm{fr}}} \nonumber\\
&{\quad +T(d_zd_h+L_Bd_h^2)} \nonumber\\
&{\quad +L_A\big[N_h(Td_hd_k+T^2d_k)+Td_hd_{\mathrm{ff}}\big]\big]\Big),}
\label{eq:complexity}
\end{align}
\looseness=1\relax {where $C_{\mathrm{ori}}$ is the orientation-network cost per sequence, $d_z=d_{sp}+d_{st}+2d_e$ is the recurrent input dimension, and $L_B$ is the number of recurrent layers. The recurrent term includes input and hidden-state projections. The attention-stack term accounts for projections, pairwise snapshot interactions, and the feedforward network in each layer~\cite{vaswani2017attention}.}

\looseness=1\relax {The embedding table stores $\mathcal{O}(U_{\max}d_e)$ parameters, where $U_{\max}$ is the number of supported target indices. Each contributor requires a conditioned prediction. Batching can exploit \ac{GPU} parallelism, with latency determined by batch size, hardware, and memory.}

{The estimator runs at the \ac{BS} using \ac{RSSI} measurements reported by the \acp{UE}. The position and orientation estimates are available for network control. The deployment latency budget includes measurement reporting and model inference, plus communication delay when estimates are returned to the devices.}

\section{Experimental Setup and Performance Evaluation}
\label{sec:experimental_setup}

\begin{table}[!t]
\centering
\caption{Key Experimental and System Parameters}
\manutablefont
\setlength{\tabcolsep}{4pt}
\begin{tabularx}{\columnwidth}{@{}l>{\raggedright\arraybackslash}X@{}}
\toprule
\textbf{Parameter} & \textbf{Value} \\
\midrule
\multicolumn{2}{c}{\textit{Experimental Setup Parameters}} \\
Carrier Frequency, $f_c$ & 28~GHz \\
System Bandwidth, $B$ & 40~MHz \\
RIS Elements, $N_R$ & 400 (20$\times$20) \\
UE Beam Patterns, $N_w$ & 60 \\
RIS Configurations, $N_{\phi}$ & 15 \\
BS-to-RIS distance, $d_{\mathrm{BS}}$ & 3~m \\
BS--RIS angle, $\vartheta_{\mathrm{BS}}$ & $-30^{\circ}$ \\
UE-to-RIS distance, $d_{\mathrm{UE}}$ & 3--7~m \\
UE--RIS angular range, $\vartheta_m$ & $0^{\circ}$--$70^{\circ}$ \\
Path loss exponent, $\gamma_{\mathrm{pl}}$ & 2.5\\
\midrule
\multicolumn{2}{c}{\textit{Model and Training Parameters}} \\
Sequence Length, $T$ & 10 \\
Number of Expert CNNs, $N_e$ & 4 \\
{Spatial Channels, $d_F$} & \leavevmode{48} \\
{BiLSTM hidden units per direction} & 128 \\
{BiLSTM layers, $L_B$} & \leavevmode{2} \\
{Temporal attention layers, $L_A$} & \leavevmode{2} \\
{Attention feedforward width, $d_{\mathrm{ff}}$} & \leavevmode{512} \\
Spatial Feature Dim., $d_{sp}$ & 128 \\
Stat. Feature Dim., $d_{st}$ & 32 \\
Gating Hidden Dim., $d_g$ & 64 \\
Batch Size & 32 \\
Initial Learning Rate & $5\times10^{-4}$ \\
Validation Strategy & 5-Fold Cross-Validation \\
Loss Weights, $\lambda_{\mathrm{ori}} / \lambda_{\mathrm{loc}}$ & 0.2 / 0.8 \\
Huber Threshold, $\delta_H$ & 1.0~m \\
CNN Kernel Sizes, $k_s$ & \{3, 5, 7\} \\
Embedding Dim., $d_e$ & 32 \\
\bottomrule
\end{tabularx}
\label{tab:sys_params}
\end{table}

\subsection{Testbed, Data Acquisition, and Implementation}

\subsubsection{Experimental Testbed and Environment Description}
\label{subsec:testbed}

The experimental platform in Fig.~\ref{fig:experimental_setup} was established in an indoor laboratory under enforced \ac{NLoS} conditions~\cite{advani2025dsog,kanaan2025multimodal}. It operates at $f_c=28$~GHz. The \ac{BS} consists of a signal generator and {an EVK02004 phased array module with a $4\times4$ uniform planar antenna array}. The \ac{UE} uses an identical array connected to {a USRP X310 software-defined radio with UBX daughterboards} on a mobile rover.

\looseness=1\relax The programmable \ac{RIS} measures $77\times77$~mm$^2$ and contains $N_R=400$ passive elements in a $20\times20$ grid. {An ESP-WROOM-32 module controls the RIS and communicates serially with a laptop that coordinates acquisition and control signaling.} The controller selects phase shifts from a fixed codebook of $N_\phi=15$ discrete Fourier transform beamsteering configurations covering a $90^{\circ}$ sector, with an average angular resolution of approximately $6^{\circ}$. The receiver uses $N_w=60$ fixed beam directions selected from a $10^{\circ}$ angular grid spanning $\pm40^{\circ}$ in azimuth and $\pm30^{\circ}$ in elevation.

The \ac{BS} and mobile \ac{UE} share an elevation, with the \ac{RIS} mounted higher. A large obstruction blocks the direct \ac{BS}--rover path. Table~\ref{tab:sys_params} summarizes the system parameters.

\subsubsection{Data Acquisition and Dataset Characteristics}
\label{subsec:data_acquisition}

The mobile rover followed 45 trajectories at 3~m to 7~m from the \ac{RIS}, with angular positions spanning 0$^{\circ}$ to 70$^{\circ}$. Measurements were acquired at four \ac{UE} orientations relative to the \ac{RIS}: 0$^{\circ}$ (broadside), 26.6$^{\circ}$, 45$^{\circ}$, and 63.4$^{\circ}$.
At each measurement point, a complete \ac{RSSI} matrix, $\mathbf{R} \in \mathbb{R}^{60 \times 15}$, was captured by sweeping all combinations of the $N_w=60$ \ac{UE} beam patterns and $N_{\phi}=15$ \ac{RIS} configurations. The dataset comprises 1,534 measurement samples. {Ground-truth position labels are obtained from the horizontal centroid coordinates in the paired light detection and ranging (LiDAR) point cloud annotations. The ZED~2 camera in Figs.~\ref{fig:system_model} and~\ref{fig:experimental_setup} serves camera-assisted beam selection on the same platform~\cite{kanaan2025multimodal} and provides no labels or inputs here.} Table~\ref{tab:dataset_stats} gives the distribution across orientations and spatial ranges.

\begin{table}[!t]
\centering
\caption{Dataset Distribution and Coverage Characteristics}
\manutablefont
\setlength{\tabcolsep}{2.5pt}
\begin{tabularx}{\columnwidth}{@{}*{5}{Y}@{}}
\toprule
\textbf{Orientation} & \textbf{Samples} & \textbf{Trajectories} & \shortstack{\textbf{Distance}\\\textbf{Range (m)}} & \shortstack{\textbf{Angular}\\\textbf{Range ($^\circ$)}} \\
\midrule
0.0$^{\circ}$ & {399} & 10 & 3.2--6.8 & 5--65 \\
26.6$^{\circ}$ & 395 & 11 & 3.0--7.0 & 0--70 \\
45.0$^{\circ}$ & 542 & 15 & 3.1--6.9 & 2--68 \\
63.4$^{\circ}$ & {198} & 9  & 3.3--6.7 & 8--62 \\
\midrule
\textbf{Total} & \textbf{1,534} & \textbf{45} & \textbf{3.0--7.0} & \textbf{0--70} \\
\bottomrule
\end{tabularx}
\label{tab:dataset_stats}
\end{table}

\begin{figure}[!t]
\centering
\definecolor{figfivebox}{RGB}{230,243,255}
\begin{tikzpicture}
\node[anchor=south west,inner sep=0pt] (fivebase) at (0,0) {\includegraphics[width=\columnwidth]{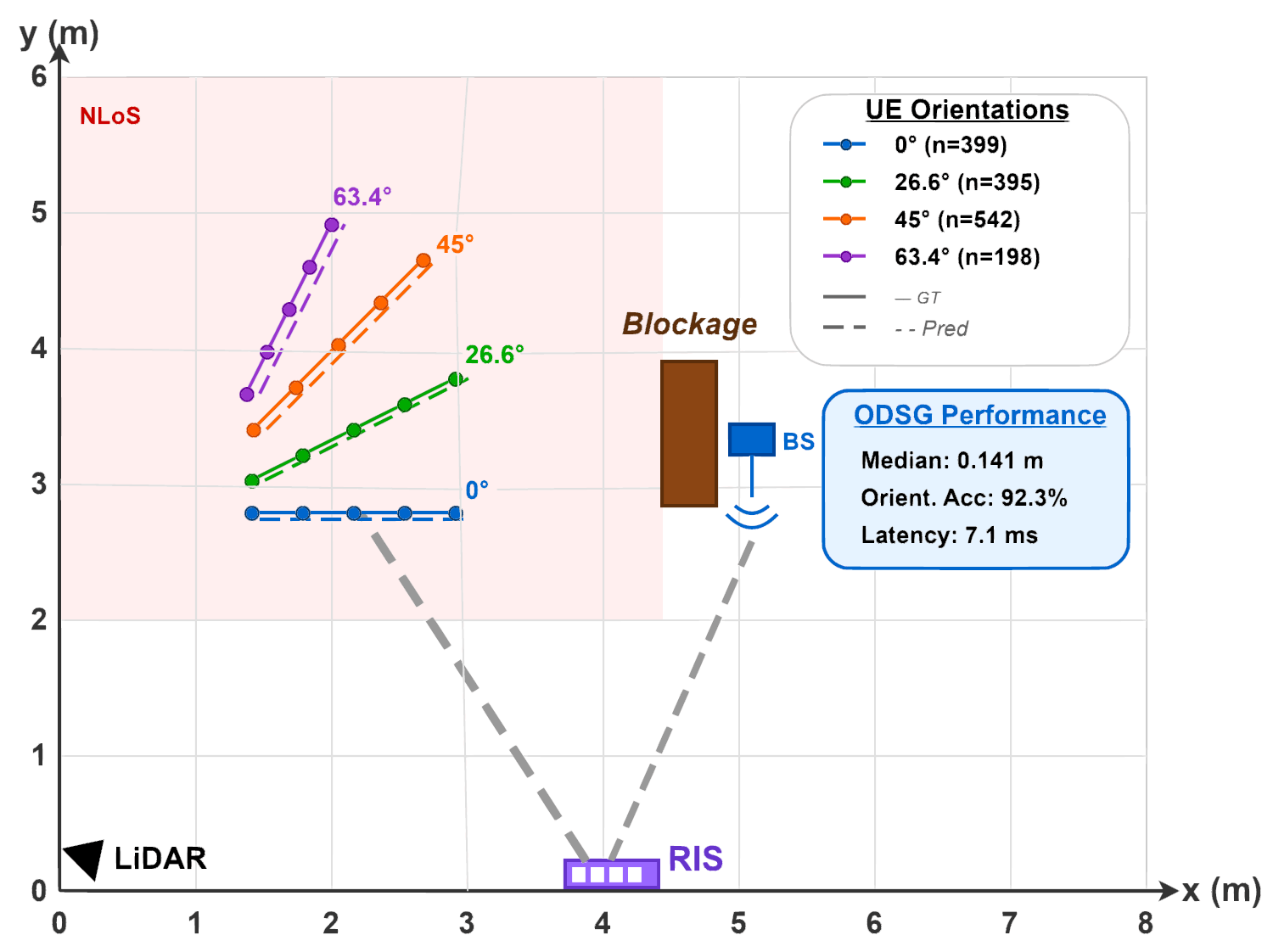}};
\begin{scope}[x={(fivebase.south east)},y={(fivebase.north west)}]
\fill[figfivebox] (0.6750,0.4198) rectangle (0.8200,0.4519);
\node[anchor=base west,inner sep=0pt,font=\usefont{T1}{phv}{b}{n}\fontsize{4.5bp}{5bp}\selectfont] at (0.6753,0.4296) {Inference: 7.1 ms};
\end{scope}
\end{tikzpicture}

\caption{{Illustrative} experimental floor plan showing ground-truth (solid) and predicted (dashed) \ac{UE} trajectories across four discrete orientations in the \ac{NLoS} region. The \ac{ODSG} framework achieves a median localization error of 0.141~m, 92.3\% orientation accuracy, and {an inference time of 7.1~ms}. }
\label{fig:gt_trajectories}
\end{figure}

{Fig.~\ref{fig:gt_trajectories} illustrates one ground-truth and predicted track per orientation, whereas every orientation spans most of the surveyed distances and angular positions (Table~\ref{tab:dataset_stats}).} Shading marks the intended \ac{NLoS} area behind the obstruction.

\subsubsection{Dataset Preprocessing and \texorpdfstring{\ac{ODSG}}{O-DSG} Implementation}
\label{subsec:preprocessing}

The raw \ac{RSSI} measurements are processed in linear power units. {Complete trajectories or acquisition groups are assigned to five outer folds before normalization, windowing, or mixture construction. Within each outer training fold, approximately 20\% of its groups are reserved for validation. Each raw measurement record belongs to exactly one of the training, validation, or test partitions within that fold. Normalization statistics and orientation pretraining use only the training partition.} Temporal sequences contain $T=10$ chronologically ordered snapshots, with 50\% overlap (stride five) between windows from the same trajectory. {Windows and mixtures are formed entirely within their assigned partition. No test trajectory or group contributes to fitting, normalization, or model selection.}

\ac{ODSG} is implemented in PyTorch with the settings in Table~\ref{tab:sys_params}. {Single-user inputs use contributor slot $u=1$. Each embedding has dimension 32, giving a joint context of dimension 64. Concatenating this context with the 128 spatial and 32 statistical features gives a recurrent input of dimension 224. The \ac{BiLSTM} has two layers with 128 hidden units per direction and concatenates the directional outputs, giving $d_h=256$. Two temporal attention layers follow, each with four heads of dimension $d_k=64$ and feedforward hidden width 512. Temporal mean pooling and a linear head map the resulting 256 features to two position coordinates.} Training uses Adam~\cite{kingma2014adam}, an initial learning rate of $5\times10^{-4}$, weight decay of $2\times10^{-5}$, and batch size 32. Early stopping monitors validation loss with a patience of 10 epochs.

{Window length controls the trade-off between temporal context, acquisition cost, and channel variation. Longer histories provide additional context, while changes in orientation can reduce its consistency with the latest snapshot. Section~\ref{subsec:acquisition_tradeoff} compares accuracy and acquisition cost across sequence lengths.}

The baselines comprise \ac{DSOG}, Random Forest, \ac{LR}, \ac{MLP}, \ac{CNN}, and \ac{NA}. {DSOG uses the preceding dual-stream recurrent architecture with supplied orientation for gating~\cite{advani2025dsog}.} NA concatenates the flattened RSSI matrices across a temporal window and processes them with a fully connected network.
{Random Forest, LR, and MLP use the endpoint snapshot, flattened in beam-major order to 900 features. LR uses ridge regression with an intercept and regularization coefficient 10. These models share the training normalization used by the other methods. The CNN encodes each of the ten snapshots with shared convolutional layers having 64, 128, and 128 channels and kernels of size five, three, and three. Unit strides and padding preserve the grid dimensions. Batch normalization and ReLU follow each convolution. Global average pooling produces 128 features per snapshot, and their chronological concatenation forms a 1,280-dimensional vector. The regression head has hidden widths 640 and 256 and two linear outputs.}

{The MLP has hidden widths 512, 512, and 256. NA has hidden widths 1,024 and 256 after its 9,000-dimensional input. Both predict two coordinates through a linear output layer. The fully connected hidden layers use ReLU, with dropout 0.2 for MLP and CNN and 0.3 for NA. These three neural baselines use coordinatewise Huber loss, summed over the two coordinates and averaged over samples. Their Adam settings are a learning rate of $5\times10^{-4}$, weight decay of $2\times10^{-5}$, and batch size 32. Training runs for at most 200 epochs with a constant learning rate. Early stopping monitors validation loss with patience ten and restores the best epoch.}

{Comparisons use the same partitions, eligible prediction endpoints, preprocessing rules, and available training measurements. DSOG additionally receives supplied orientation. Within each outer training fold, ridge regression evaluates 19 regularization coefficients spaced logarithmically from $10^{-3}$ to $10^{3}$. The other tunable models receive 20 validation trials each. Configurations are selected by validation median Euclidean position error, using the 95th-percentile error to break ties. Validation selected the same baseline settings in all five independently tuned folds. Final stochastic fits use seeds 11, 29, and 47. Ridge regression produces one deterministic final fit per fold.}
{The appendix specifies the orientation CNN and expert layers, baseline search spaces, selected sequence-model settings, and parameter and arithmetic counts.}

\subsubsection{Emulated NOMA-style Multi-User Data Generation}
\label{subsec:multi_user_synthesis}

{For each offline mixture, we draw $U$ complete windows of ten snapshots from the same data partition, using distinct raw records for each contributor. Each source retains its acquisition order. Pairing snapshots at corresponding relative indices constructs an emulated sequence from the separately measured windows. We combine the matrices in linear power using~\eqref{eq:offline_mixture} and then apply normalization with the fixed training statistics.}
A sampling constraint enforces minimum spatial separation between contributors:
\begin{equation}
\label{eq:spatial_separation}
{\|\boldsymbol{\upsilon}_i^{(t)} - \boldsymbol{\upsilon}_j^{(t)}\| \geq d_{\min},\enspace i\neq j,\enspace t=k,\ldots,k+T-1,}
\end{equation}
where $i,j\in\{1,\ldots,U\}$ and $d_{\min}=1.5$~m. {This constraint controls the synthesized geometry at every paired snapshot.} Sampled orientations belong to the four measured classes. The weighting exponent is $\gamma_{\mathrm{pl}}=2.5$. {Weights follow each contributor's horizontal distance at the current snapshot, while the index assigned at its endpoint remains fixed within the window. The balance of received powers depends jointly on these weights and the measured fingerprints.}
{Each composite window supplies $U$ instances, one for each target index under the endpoint ordering in Section~\ref{sec:system_model}. Each instance has the corresponding endpoint position and orientation as its labels. The network estimates each target directly from the common power mixture and its contributor index. Receiver noise enters through the measured source matrices.}

{Ground-truth distances determine the weights during offline synthesis, so the resulting performance characterizes the specified weighting rule.}

The training composition is 60\% single-user samples ($U=1$, no mixing), 25\% two-user samples ($U=2$), 10\% four-user samples ($U=4$), and 5\% eight-user samples ($U=8$). {These proportions define the sampling distribution of mixture windows for offline training. Averaging the loss over the targets within each window preserves its intended weight when all contributors are used for training.} {Drawing with replacement from the measured windows of each training fold yields approximately 4,500 single-user and composite training windows per fold.} {All augmented sequences derive from the original single-user measurements.}

\subsubsection{Additional Single-User Evaluation Protocol}\label{subsec:additional_protocol}

\begin{figure}[!t]
\centering
\resizebox{0.9\columnwidth}{!}{\input{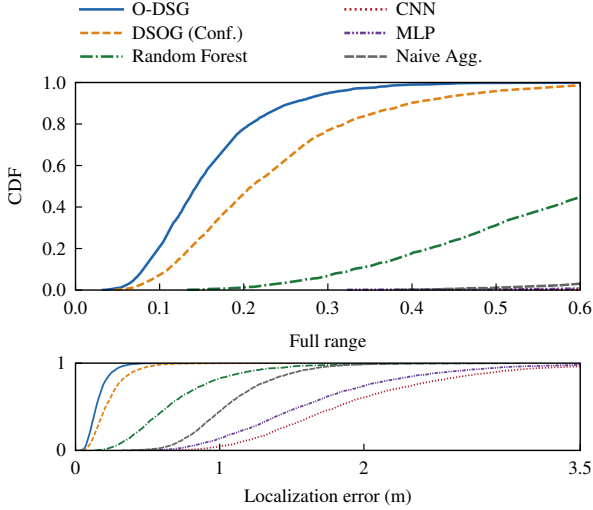}}
\caption{Cumulative distribution of localization errors {pooled over the five test folds, whereas Table~\ref{tab:main_results} averages per-fold statistics}. {The upper panel expands the 0--0.6~m range, and the lower panel retains the full error range. DSOG (Conf.) denotes the DSOG conference baseline~\cite{advani2025dsog}.}}
\label{fig:cdf_comparison}
\end{figure}

{We conduct additional experiments under single-user training to compare sequence models and assess spatial transfer, orientation transfer, and acquisition choices. The O-DSG reference uses the architecture in Section~III and is fitted separately from the mixture-trained model of Table~\ref{tab:main_results} and Figs.~\ref{fig:cdf_comparison}--\ref{fig:distance_orientation}. Tables~\ref{tab:sequence_comparison}--\ref{tab:generalization} and Fig.~\ref{fig:acquisition_tradeoff} report the additional experiments. For the sequence-model, reduced-grid, and history-length comparisons, we partition 45 groups of recorded samples into five outer folds before constructing windows. Each group remains in one partition, with approximately 20\% of the remaining groups reserved for validation. Records retain their indexed order within each group. An index gap or a position step exceeding 0.5~m starts a new segment, and each window remains within one segment. Ten-record windows with stride five yield 212 distinct test endpoints from 42 of the 45 groups across the outer folds. The sequence models and reduced-input variants use these same endpoints. Shorter inputs retain the last $T$ records of each reference window. Training records determine the scalar normalization before windows are formed.}

{Each model and condition receives 20 validation trials per fold. Selection minimizes median Euclidean position error, with the 95th percentile and parameter count used successively to break ties. Final fits use seeds 11, 29, and 47. Training uses Adam, batch size 32, at most 200 epochs, and early stopping after ten epochs without improvement in validation median error. The search considers learning rates $\{2,5,10\}\times10^{-4}$, weight decays $\{1,2,10\}\times10^{-5}$, and dropout rates $\{0.1,0.2,0.3\}$. A plateau scheduler halves the learning rate after five epochs without improvement. The sequence baselines use convolutional encoders followed by a BiLSTM, Transformer, or temporal convolutional network (TCN). DSOG receives the measured orientation label. O-DSG and the independent sequence baselines receive RSSI at inference.}

\looseness=1\relax {For these additional experiments, we compute each position-error statistic within a fold and seed, average over the three seeds, and then average equally over folds. We also report the range of the fold-averaged medians across seeds to describe initialization variability. Spatial tests use five region holdouts, and orientation tests use four angle holdouts, with the same averaging rule. Counts refer to distinct prediction endpoints and are not multiplied by the number of seeds. Inference time is the median across run-specific medians, each based on 1,000 synchronized measurements after 100 warm-up iterations at batch size one.}

\begin{table}[!t]
\caption{Localization Performance Comparison ({Original Evaluation}, Averaged over Five Folds)}
\label{tab:main_results}
\centering
\manutablefont
\setlength{\tabcolsep}{3pt}
\begin{tabular*}{\columnwidth}{@{\extracolsep{\fill}}lcccc@{}}
\toprule
\textbf{Method} & \textbf{Median} & \textbf{Mean} & \textbf{95th \%ile} & \textbf{\acs{RMSE}} \\
& \textbf{(m)} & \textbf{(m)} & \textbf{(m)} & \textbf{(m)} \\
\midrule
\acs{ODSG} & \textbf{0.141} & \textbf{0.185} & \textbf{0.437} & \textbf{0.212} \\
\acs{DSOG} & 0.190 & 0.250 & 0.590 & 0.285 \\
Random Forest & 0.750 & 0.850 & 1.450 & 0.895 \\
\ac{CNN} & 1.830 & 1.950 & 2.850 & 1.985 \\
\ac{MLP} & 1.550 & 1.650 & 2.350 & 1.685 \\
\ac{NA} & 1.030 & 1.150 & 1.850 & 1.185 \\
\ac{LR} & 2.150 & 2.320 & 3.850 & 2.425 \\
\bottomrule
\end{tabular*}
\end{table}

\subsection{Experimental Results and Performance Evaluation}

Localization performance was evaluated using the median error, \ac{MAE}, \ac{RMSE}, and 95th-percentile error. Orientation prediction was evaluated using classification accuracy and angular \ac{MAE}.

\subsubsection{Overall Localization Performance}

\begin{figure}[!t]
\centering
\includegraphics[width=0.8\columnwidth]{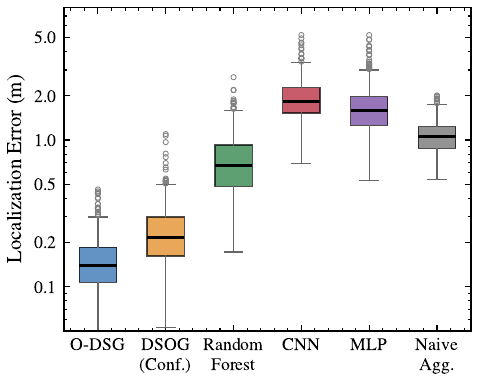}
\caption{Box plot comparison of localization errors {pooled over the five test folds}, showing medians, interquartile ranges, whiskers, and outliers. DSOG (Conf.) denotes the DSOG baseline in Table~\ref{tab:main_results}.}
\label{fig:boxplot_comparison}
\end{figure}

In Table~\ref{tab:main_results}, \ac{ODSG} achieves a median error of 0.141~m, a 25.8\% improvement over our prior \ac{DSOG} baseline~\cite{advani2025dsog} (0.190~m){, which receives the measured orientation}. {The improvement over Random Forest (0.750~m), the most accurate single-snapshot baseline, is 81.2\%.} {The DSOG, Random Forest, and CNN rows of Table~\ref{tab:main_results} are those of the evaluation in~\cite{advani2025dsog}, on the same measurements and folds.}
Fig.~\ref{fig:cdf_comparison} compares the error distributions, with an expanded low-error panel to distinguish the leading methods.
In Fig.~\ref{fig:boxplot_comparison}, \ac{ODSG} has a narrower central error distribution than the other methods.

\subsubsection{Comparison with Sequence Models}\label{subsec:sequence_comparison}

{In Table~\ref{tab:sequence_comparison}, O-DSG attains the lowest average median and 95th-percentile errors among the evaluated models. The independent sequence baselines receive the same ten-snapshot histories, so their comparison with O-DSG evaluates the complete architecture with orientation conditioning and complementary spatial and statistical features. The inference time of O-DSG for an acquired input is 7.09~ms, compared with 3.70--5.81~ms for the other models.}

\newcommand{\AdditionalOrientationTable}{
\begin{table}[!t]
\caption{{Orientation Classification in the Additional Single-User Evaluation}}
\label{tab:orientation_performance}
\centering\manutablefont
\setlength{\tabcolsep}{2.5pt}
{\begin{tabular*}{\columnwidth}{@{\extracolsep{\fill}}lcccc@{}}
\toprule
\textbf{Angle} & \textbf{Prec. (\%)} & \textbf{Recall (\%)} & \textbf{F1} & \textbf{Support} \\
\midrule
$0^{\circ}$ & 98.4 & 97.8 & 0.981 & 60 \\
$26.6^{\circ}$ & 91.1 & 91.7 & 0.914 & 56 \\
$45^{\circ}$ & 93.1 & 89.8 & 0.914 & 75 \\
$63.4^{\circ}$ & 78.3 & 87.3 & 0.824 & 21 \\
Weighted average & 92.6 & 92.3 & 0.924 & 212 \\
\bottomrule
\end{tabular*}}
\par\smallskip{\fontsize{7.5}{9}\selectfont{For each seed, predictions are pooled over the five test folds before computing class metrics. Entries are averaged over three seeds. Support counts distinct endpoints. Weighted recall equals overall accuracy.}}
\end{table}
}
\newcommand{\AdditionalGeneralizationTable}{
\begin{table}[!t]
\caption{{Generalization in the Additional Single-User Experiments}}
\label{tab:generalization}
\centering\manutablefont
\setlength{\tabcolsep}{2.5pt}
{\begin{tabular*}{\columnwidth}{@{\extracolsep{\fill}}lcccc@{}}
\toprule
\multicolumn{5}{l}{\textit{Spatial regions: median position error (m)}} \\
\textbf{Region} & \textbf{Endpoints} & \textbf{O-DSG} & \textbf{Sequence} & \\
R0 & 28 & 0.384 & 0.533 & \\
R1 & 29 & 0.218 & 0.402 & \\
R2 & 28 & 0.197 & 0.384 & \\
R3 & 32 & 0.163 & 0.270 & \\
R4 & 36 & 0.271 & 0.373 & \\
\midrule
\multicolumn{5}{l}{\textit{Omitted angles: position error and angular MAE}} \\
\textbf{Angle} & \textbf{Endpoints} & \textbf{O-DSG} & \textbf{Uncond.} & \textbf{MAE ($^\circ$)} \\
$0^{\circ}$ & 60 & 0.633 & 0.411 & 41.37 \\
$26.6^{\circ}$ & 56 & 0.297 & 0.314 & 10.36 \\
$45^{\circ}$ & 75 & 0.423 & 0.555 & 10.33 \\
$63.4^{\circ}$ & 21 & 0.575 & 0.649 & 34.78 \\
\bottomrule
\end{tabular*}}
\par\smallskip{\fontsize{7.5}{9}\selectfont{Each row averages three seeds. Position values are median errors in meters. The spatial comparator is selected on validation data within each region fold. The orientation comparator omits orientation conditioning. Angular MAE applies to O-DSG.}}
\end{table}
}
\begin{table}[!t]
\caption{{Additional Single-User Comparison with Separately Fitted Sequence Models}}
\label{tab:sequence_comparison}
\centering\manutablefont
\setlength{\tabcolsep}{2.5pt}
{\begin{tabular*}{\columnwidth}{@{\extracolsep{\fill}}lcccc@{}}
\toprule
\textbf{Model} & \textbf{Median} & \textbf{P95} & \textbf{RMSE} & \textbf{Time} \\
\midrule
O-DSG & 0.141 & 0.416 & 0.237 & 7.09 \\
DSOG & 0.190 & 0.596 & 0.309 & 5.81 \\
CNN--BiLSTM & 0.268 & 0.947 & 0.465 & 4.10 \\
CNN--Transformer & 0.284 & 0.871 & 0.463 & 4.61 \\
CNN--TCN & 0.306 & 0.928 & 0.484 & 3.70 \\
\bottomrule
\end{tabular*}}
\par\smallskip{\fontsize{7.5}{9}\selectfont{Errors are in meters and inference times in milliseconds. All models are fitted separately under the single-user protocol in Section~IV-A5. Time is the median across run medians. Median seed ranges are 0.140--0.143 (O-DSG), 0.183--0.195 (DSOG), 0.260--0.274 (BiLSTM), 0.259--0.310 (Transformer), and 0.296--0.319~m (TCN). DSOG receives supplied orientation.}}
\end{table}

\begin{figure}[!t]
\centering
\includegraphics[width=\columnwidth]{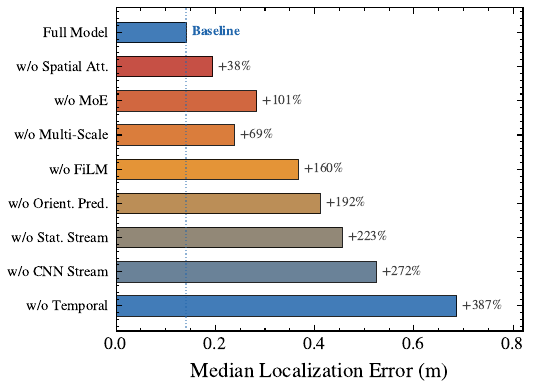}
\caption{Ablation study showing median localization error for different architectural configurations {in the original evaluation of Table~\ref{tab:main_results}}.}
\label{fig:ablation_study}
\end{figure}

\subsubsection{Ablation Study of the \texorpdfstring{\ac{ODSG}}{O-DSG}}

{Each ablation variant removes one component and is retrained with the full model's settings, so Fig.~\ref{fig:ablation_study} measures contributions within O-DSG, whereas Table~\ref{tab:sequence_comparison} compares separately tuned models.}
Removing temporal processing causes the largest degradation (387\% increase to 0.687~m). {This variant averages $\mathbf{z}^{(t)}$ over the ten snapshots before the linear head, so the comparison measures the contribution of the complete temporal block, comprising the \ac{BiLSTM} and self-attention.} Removing the \ac{CNN} stream increases the median error by 272\% (to 0.524~m), while removing the statistical aggregation stream increases it by 223\% (to 0.456~m). These results support the use of complementary spatial and statistical descriptors.
Removing the orientation prediction module increases the median error by 192\% (to 0.412~m). Removing \ac{FiLM} increases it by 160\% (to 0.367~m).
Removing the \ac{MoE} increases the median error by 101\% (to 0.283~m), while removing the multi-scale \ac{CNN} increases it by 69\% (to 0.238~m). Removing spatial attention increases it by 38\% (to 0.195~m). {Together with the \ac{FiLM} ablation, these comparisons show that orientation-conditioned feature modulation and expert weighting each reduce the error.}

\begin{figure*}[!t]
\centering
\resizebox{0.9\textwidth}{!}{\input{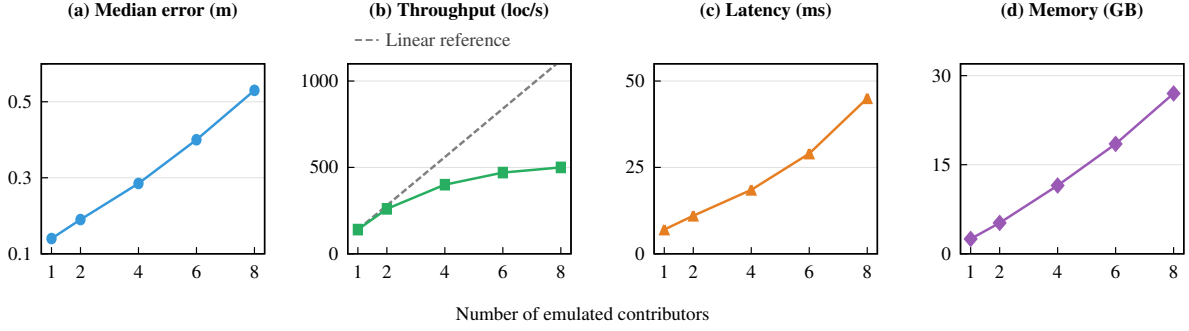}}
\caption{{Offline multi-user emulation: (a) localization error, (b) inference throughput, (c) processing latency, and (d) combined peak host and \ac{GPU} memory used by the emulation versus the number of emulated contributors. The $U$ predictions of each composite window run as one batch for throughput and sequentially at batch size one for latency. The dashed line in panel (b) is a linear scaling reference. Deployment latency additionally includes acquisition and communication.}}
\label{fig:multiuser_scalability}
\end{figure*}

\subsubsection{Emulated Multi-User Evaluation and Computational Scaling}\label{subsec:emulated_scaling}

In Fig.~\ref{fig:multiuser_scalability}(a), median error increases from 0.141~m for a single user to approximately 0.53~m for eight emulated contributors.
{The error at $U=6$, a mixture size absent from the training composition, lies between those at $U=4$ and $U=8$.} Inference throughput in Fig.~\ref{fig:multiuser_scalability}(b) increases from approximately 140 to 500 localizations per second as $U$ increases from 1 to 8, a $3.6\times$ increase. {Throughput counts target estimates when the $U$ predictions of each composite window are processed as one batch.}
Processing latency in Fig.~\ref{fig:multiuser_scalability}(c) increases from approximately 7~ms to 45~ms. {It measures completion of the $U$ predictions of one composite window, computed sequentially at batch size one as in Section~\ref{subsec:additional_protocol}. The target-independent features of the window are computed once and shared by the $U$ predictions, and batched processing in panel (b) completes more estimates per second.} {Combined peak host and \ac{GPU} memory} in panel (d) increases from 2.5~GB to 27~GB{, and the growth comes from the host-side buffers of the emulation}.

\begin{figure}[!t]
\centering
\resizebox{0.9\columnwidth}{!}{\input{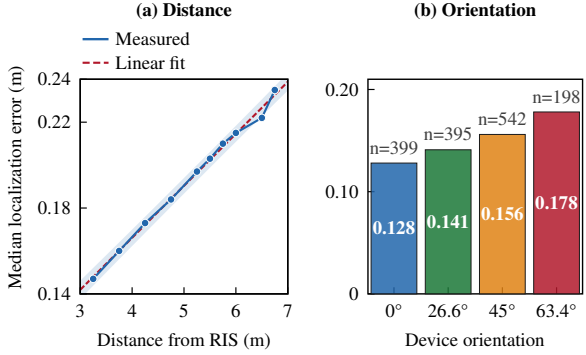}}
\caption{Performance analysis: (a) median error versus distance from \ac{RIS} showing gradual degradation with range, and (b) median error versus device orientation{, with $n$ giving the samples per orientation in Table~\ref{tab:dataset_stats}}. {Medians are averaged over the five test folds.}}
\label{fig:distance_orientation}
\end{figure}

\subsubsection{Distance, Orientation, and Prediction Analysis}

The distance analysis in Fig.~\ref{fig:distance_orientation}(a) shows median error increasing from 0.147~m at 3.25~m to 0.235~m at 6.75~m, a 60\% degradation over the 3.5~m span. The error grows approximately linearly across the measured distance bins. {The shaded band is the 95\% prediction band of the fitted line.}
The orientation analysis in Fig.~\ref{fig:distance_orientation}(b) gives the lowest median error, 0.128~m, at broadside orientation ($0^{\circ}$). The error increases to 0.178~m at $63.4^{\circ}$. {Table~\ref{tab:orientation_performance} gives the class-level orientation results on the 212 endpoints of the additional single-user evaluation. Pooling predictions over test folds for each seed and averaging across seeds gives 92.3\% accuracy and a weighted F1-score of 0.924. The $63.4^{\circ}$ class has the lowest F1-score, 0.824, and the fewest eligible endpoints.}

\AdditionalOrientationTable

{The expected angle varies continuously with the class probabilities, so it interpolates between the trained angles within $[0^{\circ},63.4^{\circ}]$. Section~\ref{subsec:generalization} tests this interpolation at omitted angles, and full azimuth coverage would add classes and a circular representation that handles angular wraparound.}

\subsubsection{Spatial and Orientation Generalization}\label{subsec:generalization}

{We assess localization at positions excluded from fitting by withholding five contiguous spatial regions in turn. To define the regions, we project the recorded coordinates $(x,y)$, in meters, onto a fixed axis, giving $c=x\cos120^{\circ}+y\sin120^{\circ}$. The projection direction was selected from coordinate coverage before fitting to retain complete temporal windows in each region. The 20th, 40th, 60th, and 80th empirical percentiles of $c$ define the boundaries between parallel spatial bands. This protocol assigns records by spatial location, separately from the group partition above. Non-test records within 0.25~m of a held-out boundary are excluded from fitting, and every retained window lies entirely within one partition. The test contains 153 endpoints across the five regions. In each spatial fold, validation performance selects the sequence comparator from the three independently tuned families, which, like O-DSG, use only RSSI at inference.}

{Table~\ref{tab:generalization} shows lower median error for O-DSG in each held-out region. The average region median is 0.247~m, compared with 0.392~m for the selected sequence comparator. Their average 95th-percentile errors are 0.553 and 0.997~m.}

{For orientation transfer, we omit each angle from training, validation, orientation pretraining, and the classifier's output support. The remaining class angles define the expected-angle estimate. We compare O-DSG with a model that omits orientation conditioning. At the interior angles $26.6^{\circ}$ and $45^{\circ}$, O-DSG gives median position errors of 0.297 and 0.423~m, compared with 0.314 and 0.555~m for the comparator. The corresponding angular MAEs are $10.36^{\circ}$ and $10.33^{\circ}$. O-DSG also improves the position median at $63.4^{\circ}$, while the comparator has lower error at $0^{\circ}$. Averaging equally over all four holdouts gives 0.482~m for both models. Angular error is larger at the endpoint holdouts because the expected angle lies within the retained class range. At both interior angles, conditioning on an angle interpolated from the retained classes lowers the position error without training data at the omitted angle.}

\AdditionalGeneralizationTable

\subsubsection{Statistical Significance and Computational Efficiency}

{For the additional comparison in Table~\ref{tab:sequence_comparison}, we pair O-DSG and DSOG within the 42 acquisition groups that produce eligible test windows. Each paired value is a group's median position error averaged over the three training seeds. The exact two-sided Wilcoxon signed-rank test gives $p=0.00145$ for this comparison. To quantify uncertainty in the average of fold medians, we perform 1,000 paired bootstrap resamples of complete acquisition groups within each test fold, retaining all windows and seed-specific predictions of each sampled group. The resulting 95\% percentile interval for the reduction in average median error is [0.019, 0.075]~m, with a point estimate of 0.049~m.}
\Ac{ODSG} uses {2.884M trainable parameters, or 11.5~MB in 32-bit floating point, compared with 1.6M for DSOG,} and reduces the median error by 25.8\% in Tables~\ref{tab:main_results} and~\ref{tab:sequence_comparison}. The inference times of \ac{ODSG} and \ac{DSOG} are 7.1 and 5.8~ms. {The \ac{ODSG} inference result on an NVIDIA RTX 3090 workstation applies to an already acquired input sequence.}
Table~\ref{tab:computational_metrics} compares {model capacity, measured inference time, and memory use. These measurements complement the operation-count analysis in Section~\ref{subsec:complexity}.}
{The single-target neural inference timings in Table~\ref{tab:computational_metrics} use batch size one, 100 warm-up iterations, and 1,000 timed measurements with accelerator synchronization. Batched throughput is benchmarked separately, as described for Fig.~\ref{fig:multiuser_scalability}.}

\begin{table}[!t]
\caption{{Model Capacity and Measured Resource Use}}
\label{tab:computational_metrics}
\centering
\manutablefont
\setlength{\tabcolsep}{3pt}
\begin{tabular*}{\columnwidth}{@{\extracolsep{\fill}}lccc@{}}
\toprule
\textbf{Method} & {\textbf{Capacity}} & \textbf{Inference} & \textbf{Memory} \\
& & \textbf{(ms)} & \textbf{(GB)} \\
\midrule
\acs{ODSG} & \leavevmode{2.884M} & 7.1 & 2.5 \\
\acs{DSOG} & 1.6M & 5.8 & 2.0 \\
\ac{CNN} & 1.2M & 4.5 & 1.5 \\
\ac{MLP} & {0.86M} & 3.2 & 1.2 \\
Random Forest & {100 trees} & 2.0 & 0.2 \\
\bottomrule
\end{tabular*}
\par\smallskip{\fontsize{7.5}{9}\selectfont{M denotes one million trainable parameters. Neural inference uses the RTX 3090; Random Forest uses all available CPU processors. {Memory is the combined peak host and \ac{GPU} memory of the inference process, including the framework runtime.}}}
\end{table}
{For sequential acquisition, let $\tau_{\mathrm{sw}}$ denote the duration of one complete snapshot sweep,}
\begin{align}
{\tau_{\mathrm{sw}}=N_wN_\phi T_p+\tau_{\mathrm{ctrl}}+\tau_{\mathrm{sched}},}
\label{eq:sweep_latency}
\end{align}
{where $\tau_{\mathrm{ctrl}}$ covers beam and RIS switching and settling outside the observation periods, and $\tau_{\mathrm{sched}}$ covers the remaining scheduling gaps. With a sliding window of the $T$ latest snapshots and sweeps repeated every $\tau_{\mathrm{upd}}\geq\tau_{\mathrm{sw}}$, the first estimate follows the start of acquisition by}
\begin{align}
{\tau_{\mathrm{first}}=(T-1)\tau_{\mathrm{upd}}+\tau_{\mathrm{sw}}+\tau_{\mathrm{proc}},}
\label{eq:first_latency}
\end{align}
{where $\tau_{\mathrm{proc}}=\tau_{\mathrm{report}}+\tau_{\mathrm{pre}}+\tau_{\mathrm{inf}}+\tau_{\mathrm{queue}}$ sums reporting, preprocessing, inference, and queueing time. A new estimate then follows every $\tau_{\mathrm{upd}}$, available $\tau_{\mathrm{proc}}$ after its newest sweep ends. Only the newest snapshot is reported for each update. With the reports of $U$ \acp{UE} handled in turn, no backlog accumulates when $U(\tau_{\mathrm{report}}+\tau_{\mathrm{pre}}+\tau_{\mathrm{inf}})\leq\tau_{\mathrm{upd}}$, and $\tau_{\mathrm{queue}}\leq(U-1)(\tau_{\mathrm{report}}+\tau_{\mathrm{pre}}+\tau_{\mathrm{inf}})$. Thus, $T$ sets the latency of the first estimate and the input history, and the sweep schedule sets the update interval.}

{In a 5G New Radio deployment, the synchronization signal timing at 120~kHz subcarrier spacing provides, every $T_{\mathrm{SS}}$, a burst of up to ${N_{\mathrm{SS}}=64}$ four-symbol blocks within 5~ms~\cite{giordani2019beam}. With one RIS configuration per burst and the $N_w\leq N_{\mathrm{SS}}$ receive beams swept across its blocks, $\tau_{\mathrm{upd}}\geq N_\phi T_{\mathrm{SS}}$ and $\tau_{\mathrm{sw}}=(N_\phi-1)T_{\mathrm{SS}}+T_{\mathrm{last}}$, where ${T_{\mathrm{last}}\leq5}$~ms spans the final burst. For $T_{\mathrm{SS}}=20$~ms, the periodicity assumed at initial cell selection, ${\tau_{\mathrm{sw}}=284}$~ms, of which the observations occupy 32~ms, and sweeps repeat every ${\tau_{\mathrm{upd}}=300}$~ms. All \acp{UE} measure the same bursts, so $\tau_{\mathrm{upd}}$ does not depend on their number. For $T=10$, \eqref{eq:first_latency} gives 2.98~s plus $\tau_{\mathrm{proc}}$. Assuming 16 bits per value and a 10~Mb/s uplink, each 900-value report, sent as uplink data, carries 14.4~kbit, adds 48~kb/s per \ac{UE}, and takes 1.44~ms. With $\tau_{\mathrm{inf}}=7.1$~ms and negligible preprocessing, $\tau_{\mathrm{proc}}\approx8.5$~ms for one \ac{UE}, plus uplink access delay.}

\subsubsection{Measurement Grid and Sequence Length}\label{subsec:acquisition_tradeoff}

\begin{figure}[!t]
\centering
\resizebox{0.9\columnwidth}{!}{\input{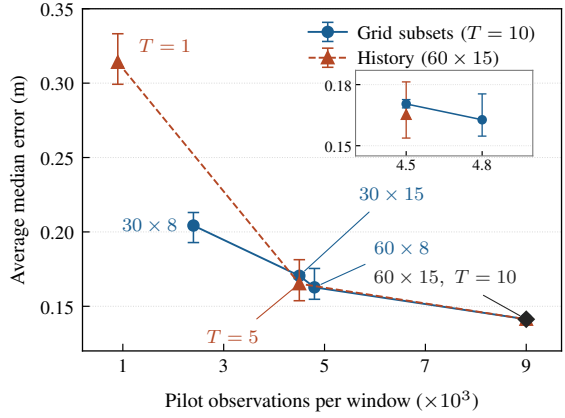}}
\caption{{Localization accuracy versus acquisition requirements under single-user training. Circles vary the measured grid at $T=10$, and triangles vary history length at the full grid. The diamond marks their shared reference. The inset expands the intermediate observation budgets. Error bars span the minimum and maximum fold-averaged medians across three seeds. Lines connect the evaluated configurations.}}
\label{fig:acquisition_tradeoff}
\end{figure}

{Fig.~\ref{fig:acquisition_tradeoff} compares acquisition choices on the 212 matched endpoints. Each model is retrained for its retained input dimensions. The reduced beam set contains indices $\{0,2,\ldots,28,31,33,\ldots,59\}$, and the reduced RIS set contains $\{0,2,\ldots,14\}$, using zero-based indices of the measured grids. These fixed subsets retain 450, 480, or 240 observations per snapshot, compared with 900 for the full grid.}

{For $T=10$, the $30\times15$, $60\times8$, and $30\times8$ grids yield average median errors of 0.171, 0.163, and 0.204~m, compared with 0.141~m for $60\times15$. Retaining eight RIS configurations reduces the observations per sweep from 900 to 480 and, in the New Radio burst schedule, $\tau_{\mathrm{upd}}$ from 300 to 160~ms, with an increase of approximately 2~cm in average median error. The $60\times8$ grid has an average 95th-percentile error of 0.497~m, compared with 0.416~m for the full grid. The $60\times8$ and $30\times15$ grids remain below the 0.190~m average median error of full-grid DSOG with supplied orientation, and the $30\times8$ grid, with 240 of the 900 observations, remains below the full-grid sequence models in Table~\ref{tab:sequence_comparison}.}

{At the full grid, $T=1$, $5$, and $10$ give average median errors of 0.314, 0.165, and 0.141~m, and average 95th-percentile errors of 0.981, 0.544, and 0.416~m. The temporal modules remain in the $T=1$ model, so this comparison measures the effect of available history separately from removing the temporal block. Ten snapshots provide the lowest errors among the tested lengths and support the reference setting. In the New Radio burst schedule, five snapshots shorten the acquisition before the first estimate from 2.98 to 1.48~s, with a 0.024~m increase in average median error. The corresponding inference times for one, five, and ten snapshots are 3.91, 5.19, and 7.09~ms.}

{Hierarchical or coded beam training offers related approaches to reducing search overhead~\cite{chen2025codedbeam}.}

{In deployment, acquisition rather than inference dominates the cost of accuracy. Relative to \ac{DSOG}, \ac{ODSG} lowers the median error by 25.8\% while its inference time rises from 5.8 to 7.1~ms, well below $\tau_{\mathrm{upd}}=300$~ms. The three sequence models in Table~\ref{tab:sequence_comparison} save at most 3.4~ms at median errors of 0.268~m or more, and the \ac{CNN}, \ac{MLP}, and Random Forest in Table~\ref{tab:computational_metrics} save at most 5.1~ms at median errors of 0.750~m or more in Table~\ref{tab:main_results}. The grid and history length therefore set the practical trade-off. With eight \ac{RIS} configurations, for example, the average median error is 0.163~m and the New Radio update interval is 160~ms.}

\subsubsection{Discussion, Implications, and Limitations}

{Section~\ref{subsec:signal_model} shows that the measured single-user fingerprints have the same mean as those of \acp{UE} served simultaneously by downlink \ac{NOMA} with a common precoder and \ac{RIS} state, for unchanged propagation channels. The offline mixture results, obtained from separately measured fingerprints, characterize an emulated task in which the contributions of several \acp{UE} overlap in one observation, as at a common receiver of simultaneous transmissions. Concurrent operation with shared observations also requires coordinated acquisition, contributor association, receiver calibration, and scheduling.}

{The orientation effect that O-DSG models is a property of the device. Rotating the \ac{UE} shifts the arrival angles in its local array coordinates (Section~\ref{subsec:signal_model}), so the power in each beam and \ac{RIS} configuration depends on orientation for any \ac{BS} and \ac{RIS} placement. With each orientation omitted from training in turn, conditioning lowers the error at three of the four, and both models average 0.482~m (Table~\ref{tab:generalization}). The exception is $0^{\circ}$, from which the expected angle stays at least $26.6^{\circ}$ away, the largest offset of any holdout. Between trained angles, the expected angle recovers the omitted orientation with an angular \ac{MAE} near $10^{\circ}$, and conditioning lowers the error at both interior holdouts. From received power alone under fixed codebooks, without \ac{CSI}, phase, or an orientation sensor, O-DSG is more accurate than DSOG supplied with the measured orientation (Table~\ref{tab:sequence_comparison}). All compared models share the data partitions and tuning budget. Accuracy follows structure rather than size, since O-DSG, with 2.884M parameters, and CNN--BiLSTM, with up to 2.462M, are both more accurate than NA with 9.48M (Tables~\ref{tab:main_results} and~\ref{tab:sequence_comparison}).}

{The orientation network, spatial stream, and experts end in global average pooling and accept grids of other sizes, so only the first statistical layer, with 19,968 of the 2.884M parameters, depends on the codebook sizes through its input dimension $N_w+N_\phi+2$. Section~\ref{subsec:acquisition_tradeoff} retrains the same architecture on three reduced grids.}

{As in any fingerprinting method, the trained weights belong to the surveyed site, so a new site needs a labeled survey and retraining, whereas the power-only input, the orientation-conditioned architecture, and the training procedure carry over. A survey of the present size, 1,534 sweeps along 45 trajectories, supports the reported accuracy, and with sweeps reduced to the $60\times8$ or $30\times15$ grid, O-DSG remains more accurate than full-grid DSOG. Within the surveyed area, O-DSG also has lower median error than the selected sequence comparator in every spatial region excluded from fitting. Transfer learning can start from the trained spatial stream and experts and update the orientation and statistical branches, with retraining cost set by the updated layers, the target survey, the optimization budget, and the hardware. Codeword-independent RIS representations~\cite{luo2023codeword} and Wi-Fi fingerprint adaptation across devices~\cite{li2024itoloc} provide related directions. Future work will apply O-DSG beyond the present room and placement, with other RIS codebooks, carrier frequencies, and hardware.}

{RSSI acquisition uses power measurements with coordinated pilot, beam, and RIS schedules. Configuration identifiers, timestamps, and switching and settling intervals preserve the correspondence between observations and grid entries. Consistent transmit power, receiver gain, and calibrated reflection patterns are needed between collection and operation~\cite{etsi2025ris004}. Each measured snapshot was acquired with the rover stationary, matching the quasi-static channel of~\eqref{eq:rssi_matrix}, and consecutive evaluated snapshots lie a median of 0.127~m apart. A \ac{UE} moving at speed $v$ reproduces the measured inputs when the spacing $v\tau_{\mathrm{upd}}$ of its snapshots is comparable to 0.127~m and its displacement $v\tau_{\mathrm{sw}}$ during each sweep is small relative to that spacing, which requires $\tau_{\mathrm{sw}}\ll\tau_{\mathrm{upd}}$. For $\tau_{\mathrm{upd}}=300$~ms, the 0.127~m spacing corresponds to $v\approx0.42$~m/s, at which a \ac{UE} moves 0.12~m during the 284~ms sweep of the 20~ms periodicity. With $T_{\mathrm{SS}}=5$~ms~\cite{giordani2019beam} and the same update interval, $\tau_{\mathrm{sw}}\approx74$~ms limits the displacement of a moving \ac{UE} to about 0.03~m and leaves about 0.5~ms between bursts for \ac{RIS} reconfiguration.}

{At the \ac{BS}, each update recomputes the ten-snapshot window with 4.283~giga multiply--accumulate operations (GMACs, Table~\ref{tab:baseline_arithmetic}), about 14.3~GMAC/s per \ac{UE} at $\tau_{\mathrm{upd}}=300$~ms. The target platform sets the corresponding execution time, peak memory, and energy, and quantization or pruning could reduce them.}

\subsubsection{Relationship to Adaptive RIS Learning}
\label{subsec:drl_tradeoffs}
{The supervised estimator learns from labeled positions and orientations and processes prescribed sweeps. RIS controllers based on DDPG or DDQN learn through action--reward interactions, exploration, replay, and policy or value updates~\cite{chou2025drl,wang2025heuristicdrl}. DDPG supports continuous controls, whereas DDQN uses discrete actions. After training, a policy adjusts its actions in response to incoming observations. Maintaining performance beyond its training distribution may require further updates. Training costs therefore depend on the labeled calibration data needed by supervised learning and the interactions and feedback needed by DRL.}

{At deployment, both approaches incur observation and control costs beyond network inference. Online DRL exploration, when enabled, adds interactions with the controlled system.}

{Combining adaptive RIS control with localization would require a reward tied to localization and an estimator compatible with the chosen configurations. Chou et al.~\cite{chou2025drl} use information about user presence and rate feedback, while Wang et al.~\cite{wang2025heuristicdrl} assume channel information.}

\section{Conclusion}

This paper presented a dual-stream gated localization framework that uses orientation inferred from RSSI to condition feature modulation and soft expert gating. Recurrent processing and temporal attention combine information across snapshots. Measurements at 28~GHz through a 400-element \ac{RIS} under \ac{NLoS} conditions show a median error of 0.141~m from received power alone, and ablations support the complementary spatial, statistical, and temporal stages. {Additional comparisons with sequence models show gains in median and tail error on matched inputs. In every held-out spatial region, O-DSG has lower median error than the selected sequence model, 0.247 versus 0.392~m on average, and orientation conditioning lowers the error at three of the four omitted orientations. With eight of the 15 \ac{RIS} configurations, O-DSG remains more accurate than full-grid DSOG while the New Radio update interval falls from 300 to 160~ms.} {Under downlink \ac{NOMA} with a common precoder and \ac{RIS} state, each \ac{UE}'s mean fingerprint equals its single-user value for unchanged propagation channels, and one common sweep yields the fingerprints of any number of \acp{UE}. The offline mixture study, built from separately measured fingerprints, tests localization of requested contributors from a shared composite and measures computational scaling with the number of contributors.}

\appendix[Network Implementation and Model Selection]
\label{app:baselines}

{This appendix supplements Section~IV-A.}

\subsection{Orientation Network and Expert Layers}

{For the full measurement grid and all four orientation classes, the orientation CNN processes the latest normalized snapshot as a single-channel $60\times15$ grid. Its three convolutions have 32, 64, and 64 output channels. Each uses a $3\times3$ kernel, unit stride, one-cell zero padding, and a bias, followed by batch normalization and ReLU. A $2\times2$ max-pooling layer with stride two follows each of the second and third convolutional blocks. Pooling rounds output dimensions down, giving grids of $30\times7$ and $15\times3$, respectively. Global average pooling produces 64 features. Concatenating the shared 32-dimensional contributor embedding gives 96 inputs to a classification head with a 64-unit ReLU hidden layer and four output logits. Softmax yields the class posterior.}

{Each of the four experts receives the FiLM-modulated tensor with $d_F=48$ channels on the $60\times15$ grid. A $3\times3$ convolution preserves all 48 channels and both grid dimensions using unit stride, one-cell zero padding, and a bias. Batch normalization and ReLU precede global average pooling. The resulting 48 features pass through a 512-unit fully connected layer, ReLU, dropout with probability 0.2, and a linear projection to 128 features. Each expert has its own parameters, shared across all time steps. Dense layers in the orientation head and experts include biases.}

{The 64-dimensional joint context produces 48 FiLM scales and 48 shifts. The gate concatenates this context with 48 pooled spatial features and maps the resulting 112 inputs through 64 ReLU units to four Softmax weights. The weighted expert output has 128 dimensions. Together with 32 statistical features and the 64-dimensional context, it forms the 224-dimensional temporal input specified in Section~IV-A.}

\subsection{Statistical and Temporal Layers}

{The statistical MLP maps the 77 descriptors to 256 features through an affine layer and ReLU. A residual branch applies two 256-unit affine layers with ReLU between them. The residual sum passes through ReLU. A final affine projection produces 32 statistical features. All affine layers include biases. This branch contains 159,776 trainable parameters.}

{The two BiLSTM layers each have 128 hidden units per direction. Their input widths are 224 and 256, respectively, and together they contain 757,760 trainable parameters. Each of the two subsequent attention layers uses four 64-dimensional heads and a feedforward network with widths 256, 512, and 256. Layer normalization acts across the 256 features of each snapshot. Attention and feedforward projections include biases. Temporal mean pooling precedes the final affine map from 256 features to two coordinates.}

\subsection{Original Baselines}

{Random Forest uses one multi-output estimator with 100 trees, bootstrap sampling, squared-error splitting, unrestricted depth, minimum split size two, and minimum leaf size two. Each split considers 297 of the 900 features, corresponding to a feature fraction of 0.33. The search considers tree counts $\{100,200,300,500\}$, maximum depths $\{10,20,30,\text{unrestricted}\}$, minimum leaf sizes $\{1,2,4\}$, and the square-root feature rule or fractions $\{0.33,1.0\}$. LR uses ridge regression with an intercept, automatic solver selection, and no feature expansion. Its candidate coefficients are $\{10^{-3+j/3}:j=0,\ldots,18\}$, with 10 selected.}

{The MLP search considers two to four hidden layers with widths $\{256,512,768\}$. CNN varies its initial channel width over $\{32,64,96\}$, initial kernel size over $\{3,5\}$, convolutional depth from two to four layers, and first regression-layer width over $\{256,512,640\}$. NA searches first and second hidden widths $\{512,1024\}$ and $\{128,256\}$, respectively. All three searches use learning rates $\{2,5,10\}\times10^{-4}$ and weight decays $\{1,2,10\}\times10^{-5}$. Dropout candidates are $\{0.1,0.2,0.3\}$ for MLP and CNN and $\{0.2,0.3,0.5\}$ for NA. Their coordinatewise Huber loss uses a threshold of 1~m, and Adam uses $(\beta_1,\beta_2)=(0.9,0.999)$. MLP and NA have no normalization layers. CNN uses batch normalization in its convolutional encoder.}

\subsection{Additional Sequence Baselines}

{The shared snapshot encoder has three $3\times3$ convolutions with 32, 64, and 128 channels, the last two with stride two. Batch normalization and ReLU follow each convolution, and global average pooling yields the snapshot features. Position regression uses the final recurrent, token, or temporal-convolution output. The Transformer uses learned positional embeddings. TCN dilation starts at one and doubles per residual level.}

{BiLSTM and Transformer searches use widths $\{128,\allowbreak 192,\allowbreak 256\}$ and one to three layers. The TCN search uses widths $\{96,128,192\}$, three to five residual levels, and kernel sizes two or three. All three searches consider CNN channel multipliers one and two. The learning-rate, weight-decay, dropout, stopping, and scheduling rules are specified in Section~IV-A.}

{Tuples list selected settings in fold order 0\nobreakdash--4. CNN--BiLSTM uses hidden units per direction $(192,\allowbreak 192,\allowbreak 256,\allowbreak 192,\allowbreak 192)$ and layer counts $(2,3,2,2,2)$. CNN--Transformer uses width 192, two layers, four heads, and feedforward width 384 in every fold. CNN--TCN uses 128 channels, kernel size three, and residual-level counts $(4,3,4,4,4)$. The selected CNN channel multiplier is one throughout.}

{For CNN--BiLSTM, CNN--Transformer, and CNN--TCN, respectively, learning-rate tuples in units of $10^{-4}$ are $(2,2,10,5,5)$, $(5,5,5,5,5)$, and $(10,5,5,5,5)$. Weight-decay tuples in units of $10^{-5}$ are $(2,1,10,2,1)$, $(10,1,2,2,2)$, and $(2,2,2,2,2)$. Dropout tuples are $(0.2,0.3,0.2,0.2,0.3)$, $(0.1,0.2,0.1,0.1,0.3)$, and $(0.1,0.2,0.2,0.2,0.2)$. The resulting parameter counts are 1.476--2.462M, 0.714M, and 0.391--0.490M, respectively.}

\subsection{Parameter and Operation Counts}

\looseness=1\relax {Table~\ref{tab:baseline_arithmetic} gives counts for O-DSG, the specified original neural baselines, and LR. A convolution contributes its output grid area times its kernel area and input and output channel counts. A dense layer contributes the product of its input and output widths. One multiply--accumulate (MAC) counts as two floating-point operations. Counts include convolutional, dense, recurrent, and attention matrix products per endpoint prediction. Bias additions, normalization, activation, pooling, residual additions, and elementwise feature modulation are excluded. Parameter counts include biases and trainable normalization coefficients. Ridge includes both coordinate coefficient vectors and intercepts.}

\looseness=1\relax {O-DSG's count uses the full ten-snapshot window and all four experts at each time step. Orientation inference and the FiLM context projections run once per window. The shared contributor embedding is counted once. The resulting 4.283~GMACs correspond to 8.566~GFLOPs.}

\begin{table}[!t]
\caption{{Model Parameters and Arithmetic Counts}}
\label{tab:baseline_arithmetic}
\centering
\fontsize{8}{9.5}\selectfont
\setlength{\tabcolsep}{3pt}
{\begin{tabular*}{\columnwidth}{@{\extracolsep{\fill}}lrr@{}}
\toprule
\textbf{Model} & \textbf{Parameters} & \textbf{MACs per prediction} \\
\midrule
O-DSG & $2\,884\,382$ & $4\,283\,107\,872$ \\
MLP & $855\,810$ & $854\,528$ \\
CNN & $1\,208\,194$ & $2\,006\,039\,552$ \\
NA & $9\,479\,938$ & $9\,478\,656$ \\
LR (ridge) & $1\,802$ & $1\,800$ \\
\bottomrule
\end{tabular*}}
\end{table}

\bibliographystyle{IEEEtran}
\bibliography{references}

@article{li2024indoor,
  author    = {Li, Peisen and Wu, Wei and Zhao, Zhiheng and Huang, George Q.},
  title     = {{Indoor Positioning Systems in Industry 4.0 Applications: Current Status, Opportunities, and Future Trends}},
  journal   = {Digital Engineering},
  volume    = {3},
  pages     = {100020},
  year      = {2024},
  doi       = {10.1016/j.dte.2024.100020},
  publisher = {Elsevier}
}

@book{proakis2008digital,
  author    = {Proakis, John G. and Salehi, Masoud},
  title     = {{Digital Communications}},
  edition   = {5},
  publisher = {McGraw-Hill},
  address   = {New York, NY, USA},
  year      = {2008}
}

@article{liu2021reconfigurable,
  author = {Liu, Yuanwei and Liu, Xiao and Mu, Xidong and Hou, Tianwei and Xu, Jiaqi and Di Renzo, Marco and Al-Dhahir, Naofal},
  title = {{Reconfigurable Intelligent Surfaces: Principles and Opportunities}},
  journal = {IEEE Communications Surveys \& Tutorials},
  volume = {23}, number = {3}, pages = {1546--1577}, year = {2021},
  doi = {10.1109/COMST.2021.3077737}
}

@inproceedings{rahal2023ris,
  title={{RIS}-aided positioning experiments based on {mmWave} indoor channel measurements},
  author={Rahal, Moustafa and Denis, Beno{\^\i}t and Mazloum, Taghrid and Munoz, Frederic and D'Errico, Raffaele},
  booktitle={2023 13th International Conference on Indoor Positioning and Indoor Navigation (IPIN)},
  pages={1--6},
  year={2023},
  organization={IEEE}
}

@article{dardari2021nlos,
  title={{LOS}/{NLOS} near-field localization with a large reconfigurable intelligent surface},
  author={Dardari, Davide and Decarli, Nicol{\'o} and Guerra, Anna and Guidi, Francesco},
  journal={IEEE Transactions on Wireless Communications},
  volume={21},
  number={6},
  pages={4282--4294},
  year={2022},
  doi={10.1109/TWC.2021.3128415},
  publisher={IEEE}
}

@article{zhang2021metalocalization,
  author = {Zhang, Haobo and Zhang, Hongliang and Di, Boya and Bian, Kaigui and Han, Zhu and Song, Lingyang},
  title = {{MetaLocalization: Reconfigurable Intelligent Surface Aided Multi-User Wireless Indoor Localization}},
  journal = {IEEE Transactions on Wireless Communications},
  volume = {20}, number = {12}, pages = {7743--7757}, year = {2021},
  doi = {10.1109/TWC.2021.3087354}
}

@article{keykhosravi2022ris,
  title={{RIS}-enabled {SISO} localization under user mobility and spatial-wideband effects},
  author={Keykhosravi, Kamran and Keskin, Musa Furkan and Seco-Granados, Gonzalo and Popovski, Petar and Wymeersch, Henk},
  journal={IEEE Journal of Selected Topics in Signal Processing},
  volume={16},
  number={5},
  pages={1125--1140},
  year={2022},
  publisher={IEEE}
}

@article{rizk2022robust,
  title={A robust and accurate indoor localization using learning-based fusion of {Wi-Fi RTT} and {RSSI}},
  author={Rizk, Hamada and Elmogy, Ahmed and Yamaguchi, Hirozumi},
  journal={Sensors},
  volume={22},
  number={7},
  pages={2700},
  year={2022},
  publisher={MDPI}
}

@article{burghal2020comprehensive,
  author  = {Burghal, Daoud and Ravi, Ashwin T. and Rao, Varun and Alghafis, Abdullah A. and Molisch, Andreas F.},
  title   = {{A Comprehensive Survey of Machine Learning Based Localization with Wireless Signals}},
  journal = {arXiv preprint arXiv:2012.11171},
  year    = {2020}
}

@inproceedings{aguiar2024deep,
  title={A deep learning approach in {RIS}-based indoor localization},
  author={Aguiar, Rafael A and Paulino, Nuno and Pessoa, Lu{\'\i}s M},
  booktitle={2024 Joint European Conference on Networks and Communications \& 6G Summit (EuCNC/6G Summit)},
  pages={523--528},
  year={2024},
  organization={IEEE}
}

@article{nguyen2010reconfigurable,
  author    = {Nguyen, C. and Georgiou, O. and Gradoni, G.},
  title     = {Reconfigurable Intelligent Surfaces and Machine Learning for Wireless Fingerprinting Localization},
  journal   = {arXiv preprint arXiv:2010.03251},
  year      = {2020}
}

@inproceedings{advani2025dsog,
  author    = {Advani, Varun S. and Nasser, Ahmed and Selim, Mohamed Y. and Eltawil, Ahmed M.},
  title     = {{Deep Learning Framework for RSSI-Based Indoor
Localization in RIS-Aided mmWave Systems}},
  booktitle = {Proc. IEEE Global Communications Conference (GLOBECOM)},
  pages     = {4233--4238},
  year      = {2025},
  doi       = {10.1109/GLOBECOM59602.2025.11431957}
}

@article{huang2023joint,
  title={Joint localization and environment sensing by harnessing {NLOS} components in {RIS}-aided {mmWave} communication systems},
  author={Huang, Yixuan and Yang, Jie and Tang, Wankai and Wen, Chao-Kai and Xia, Shuqiang and Jin, Shi},
  journal={IEEE Transactions on Wireless Communications},
  volume={22},
  number={12},
  pages={8797--8813},
  year={2023},
  publisher={IEEE}
}

@inproceedings{sun2024near,
  title={Near-field {RIS}-aided Localization Under Channel Non-Stationarity: A Mismatched Model Approach},
  author={Sun, Bo and Keskin, Musa Furkan and Rahal, Moustafa and Chen, Hui and Talvitie, Jukka and Wymeersch, Henk and Valkama, Mikko},
  booktitle={2024 IEEE 25th International Workshop on Signal Processing Advances in Wireless Communications (SPAWC)},
  pages={271--275},
  year={2024},
  organization={IEEE}
}

@article{jacobs1991adaptive,
  author  = {Jacobs, Robert A. and Jordan, Michael I. and Nowlan, Steven J. and Hinton, Geoffrey E.},
  title   = {{Adaptive Mixtures of Local Experts}},
  journal = {Neural Computation},
  volume  = {3},
  number  = {1},
  pages   = {79--87},
  year    = {1991}
}

@inproceedings{perez2018film,
  author    = {Perez, Ethan and Strub, Florian and De Vries, Harm and Dumoulin, Vincent and Courville, Aaron},
  title     = {{FiLM: Visual Reasoning with a General Conditioning Layer}},
  booktitle = {Proc. AAAI Conference on Artificial Intelligence},
  volume    = {32},
  number    = {1},
  year      = {2018}
}

@inproceedings{vaswani2017attention,
  author    = {Vaswani, Ashish and Shazeer, Noam and Parmar, Niki and Uszkoreit, Jakob and Jones, Llion and Gomez, Aidan N. and Kaiser, {\L}ukasz and Polosukhin, Illia},
  title     = {{Attention Is All You Need}},
  booktitle = {Advances in Neural Information Processing Systems},
  pages     = {5998--6008},
  year      = {2017}
}

@article{11175018,
  author   = {Nasser, Ahmed and Celik, Abdulkadir and Abdallah, Asmaa and Wang, Ruiqi and Yang, Yiming and Shamim, Atif and Eltawil, Ahmed M.},
  title    = {{Acing Versatility of {RIS} Partitioning: An Experimental Proof-of-Concept Campaign}},
  journal  = {IEEE Communications Standards Magazine},
  volume   = {9},
  number   = {4},
  month    = dec,
  pages    = {144--151},
  year     = {2025},
  doi      = {10.1109/MCOMSTD.2025.3608130}
}

@article{11218162,
  author   = {Nasser, Ahmed and Celik, Abdulkadir and Eltawil, Ahmed M.},
  title    = {{Multi-Agent {DRL} for {RIS} Partitioning, Beam Selection, and Power Control in {MIMO}-{NOMA} System}},
  journal  = {IEEE Open Journal of the Communications Society},
  volume   = {6},
  pages    = {9073--9089},
  year     = {2025},
  doi      = {10.1109/OJCOMS.2025.3624260}
}

@article{liu2022isac_survey,
  author  = {Liu, An and Huang, Zhe and Li, Min and Wan, Yubo and Li, Wenrui and Han, Tony Xiao and Liu, Chenchen and Du, Rui and Tan, Danny Kai Pin and Lu, Jianmin and others},
  title   = {{A Survey on Fundamental Limits of Integrated Sensing and Communication}},
  journal = {IEEE Communications Surveys \& Tutorials},
  volume  = {24},
  number  = {2},
  pages   = {994--1034},
  year    = {2022},
  doi     = {10.1109/COMST.2022.3149272}
}

@article{zafari2019indoor_survey,
  author  = {Zafari, Faheem and Gkelias, Athanasios and Leung, Kin K.},
  title   = {{A Survey of Indoor Localization Systems and Technologies}},
  journal = {IEEE Communications Surveys \& Tutorials},
  volume  = {21},
  number  = {3},
  pages   = {2568--2599},
  year    = {2019},
  doi     = {10.1109/COMST.2019.2911558}
}

@article{busari2018mmwave_survey,
  author  = {Busari, Sherif Adeshina and Huq, Kazi Mohammed Saidul and Mumtaz, Shahid and Dai, Linglong and Rodriguez, Jonathan},
  title   = {{Millimeter-Wave Massive {MIMO} Communication for Future Wireless Systems: {A} Survey}},
  journal = {IEEE Communications Surveys \& Tutorials},
  volume  = {20},
  number  = {2},
  pages   = {836--869},
  year    = {2018},
  doi     = {10.1109/COMST.2017.2787460}
}

@inproceedings{he2016deep,
  author    = {He, Kaiming and Zhang, Xiangyu and Ren, Shaoqing and Sun, Jian},
  title     = {{Deep Residual Learning for Image Recognition}},
  booktitle = {Proc. IEEE Conference on Computer Vision and Pattern Recognition (CVPR)},
  pages     = {770--778},
  year      = {2016}
}

@article{graves2005framewise,
  title={Framewise phoneme classification with bidirectional {LSTM} and other neural network architectures},
  author={Graves, Alex and Schmidhuber, J{\"u}rgen},
  journal={Neural Networks},
  volume={18},
  number={5-6},
  pages={602--610},
  year={2005},
  publisher={Elsevier}
}

@inproceedings{kingma2014adam,
  title={Adam: A method for stochastic optimization},
  author={Kingma, Diederik P and Ba, Jimmy},
  booktitle={International Conference on Learning Representations (ICLR)},
  year={2015}
}

@article{fadakar2024multi,
  author = {Fadakar, Alireza and Sabbaghian, Maryam and Wymeersch, Henk},
  title = {{Multi-RIS-Assisted 3D Localization and Synchronization via Deep Learning}},
  journal = {IEEE Open Journal of the Communications Society},
  volume = {5}, pages = {3299--3314}, year = {2024},
  doi = {10.1109/OJCOMS.2024.3399605}
}

@article{shahmansoori2018posnavmmwave,
  author = {Shahmansoori, Arash and Garcia, Gabriel E. and Destino, Giuseppe and Seco-Granados, Gonzalo and Wymeersch, Henk},
  title = {{Position and Orientation Estimation Through Millimeter-Wave MIMO in 5G Systems}},
  journal = {IEEE Transactions on Wireless Communications},
  volume = {17}, number = {3}, pages = {1822--1835}, year = {2018},
  doi = {10.1109/TWC.2017.2785788}
}

@article{palacios2019mmwave,
  author  = {Palacios, Joan and Bielsa, Guillermo and Casari, Paolo and Widmer, Joerg},
  title   = {{Single- and Multiple-Access Point Indoor Localization for Millimeter-Wave Networks}},
  journal = {IEEE Transactions on Wireless Communications},
  volume  = {18},
  number  = {3},
  pages   = {1927--1942},
  year    = {2019},
  doi     = {10.1109/TWC.2019.2899313}
}

@article{wang2018noma,
  title   = {{Non-Orthogonal Multiple Access for 5G: Solutions, Challenges, Opportunities, and Future Research Trends}},
  author  = {Dai, Linglong and Wang, Bichai and Yuan, Yifei and Han, Shuangfeng and I, Chih-Lin and Wang, Zhaocheng},
  journal = {IEEE Communications Magazine},
  volume  = {53},
  number  = {9},
  pages   = {74--81},
  year    = {2015},
  month   = sep,
  doi     = {10.1109/MCOM.2015.7263349}
}

@article{chou2025drl,
  author  = {Chou, Po-Heng and Zheng, Bo-Ren and Huang, Wan-Jen and Saad, Walid and Tsao, Yu and Chang, Ronald Y.},
  title   = {{Deep Reinforcement Learning-Based Precoding for Multi-RIS-Aided Multiuser Downlink Systems With Practical Phase Shift}},
  journal = {IEEE Wireless Communications Letters},
  volume  = {14},
  number  = {1},
  pages   = {23--27},
  year    = {2025},
  month   = jan,
  doi     = {10.1109/LWC.2024.3482720}
}

@article{abeywickrama2020practical,
  author  = {Abeywickrama, Samith and Zhang, Rui and Wu, Qingqing and Yuen, Chau},
  title   = {{Intelligent Reflecting Surface: Practical Phase Shift Model and Beamforming Optimization}},
  journal = {IEEE Transactions on Communications},
  volume  = {68},
  number  = {9},
  pages   = {5849--5863},
  year    = {2020},
  month   = sep,
  doi     = {10.1109/TCOMM.2020.3001125}
}

@article{wang2025heuristicdrl,
  author  = {Wang, Wei and Li, Peizheng and Doufexi, Angela and Beach, Mark A.},
  title   = {{A Heuristic-Integrated DRL Approach for Phase Optimization in Large-Scale RISs}},
  journal = {IEEE Communications Letters},
  volume  = {29},
  number  = {7},
  pages   = {1579--1583},
  year    = {2025},
  month   = jul,
  doi     = {10.1109/LCOMM.2025.3568652}
}

@article{li2024itoloc,
  author  = {Li, Danyang and Xu, Jingao and Yang, Zheng and Tang, Chengpei},
  title   = {{Train Once, Locate Anytime for Anyone: Adversarial Learning-based Wireless Localization}},
  journal = {ACM Transactions on Sensor Networks},
  volume  = {20},
  number  = {2},
  pages   = {37:1--37:21},
  year    = {2024},
  month   = jan,
  doi     = {10.1145/3614095}
}

@techreport{etsi2025ris004,
  author      = {{ETSI}},
  title       = {{Reconfigurable Intelligent Surfaces (RIS); Implementation and Practical Considerations}},
  institution = {European Telecommunications Standards Institute},
  type        = {Group Report},
  number      = {ETSI GR RIS 004 V1.1.1},
  year        = {2025},
  month       = mar
}

@article{chen2025codedbeam,
  author  = {Chen, Yuhao and Dai, Linglong},
  title   = {{Coded Beam Training for RIS-Assisted Wireless Communications}},
  journal = {IEEE Transactions on Wireless Communications},
  volume  = {24},
  number  = {5},
  pages   = {3854--3867},
  year    = {2025},
  month   = may,
  doi     = {10.1109/TWC.2025.3535160}
}

@article{luo2023codeword,
  author = {Luo, Xuanshu and Meratnia, Nirvana},
  title = {{A Codeword-Independent Localization Technique for Reconfigurable Intelligent Surface Enhanced Environments Using Adversarial Learning}},
  journal = {Sensors}, volume = {23}, number = {2}, pages = {984}, year = {2023},
  doi = {10.3390/s23020984}
}

@inproceedings{he2016identity,
  author = {He, Kaiming and Zhang, Xiangyu and Ren, Shaoqing and Sun, Jian},
  title = {{Identity Mappings in Deep Residual Networks}},
  booktitle = {European Conference on Computer Vision (ECCV)},
  year = {2016},
  url = {https://arxiv.org/abs/1603.05027}
}

@inproceedings{kanaan2025multimodal,
  author = {Kanaan, Khalid and Nasser, Ahmed and Celik, Abdulkadir and Wang, Ruiqi and Shamim, Atif and Eltawil, Ahmed M.},
  title = {{Multimodal Sensing and DRL-Driven Beam Selection in RIS-Aided mmWave mMIMO Systems}},
  booktitle = {2025 IEEE 36th Annual International Symposium on Personal, Indoor and Mobile Radio Communications (PIMRC)},
  pages = {1--6},
  year = {2025},
  doi = {10.1109/PIMRC62392.2025.11274583}
}

@article{giordani2019beam,
  author = {Giordani, Marco and Polese, Michele and Roy, Arnab and Castor, Douglas and Zorzi, Michele},
  title = {{A Tutorial on Beam Management for 3GPP NR at mmWave Frequencies}},
  journal = {IEEE Communications Surveys \& Tutorials},
  volume = {21},
  number = {1},
  pages = {173--196},
  year = {2019},
  doi = {10.1109/COMST.2018.2869411}
}

\end{document}